\documentclass[11pt]{article}
\usepackage{graphicx}
\usepackage{booktabs}
\usepackage{xcolor} 
\usepackage{colortbl}
\usepackage{makecell}
\usepackage{rotating}
\usepackage{multirow}
\usepackage{listings}
\usepackage{textcomp}
\usepackage[most]{tcolorbox}
\usepackage{courier}
\usepackage{hyperref}
\tcbuselibrary{skins,breakable}

\usepackage[final]{acl}

\usepackage{times}
\usepackage{latexsym}
\usepackage{amsmath,amssymb}
\usepackage[T1]{fontenc}

\usepackage[utf8]{inputenc}

\usepackage{microtype}

\usepackage{inconsolata}

\usepackage{graphicx}

\title{VoxMind: An End-to-End Agentic Spoken Dialogue System}

\author{
  \textbf{Tianle Liang\textsuperscript{1,2}$^*$}\quad
  \textbf{Yifu Chen\textsuperscript{1}$^*$}\quad
  \textbf{Shengpeng Ji\textsuperscript{1}$^*$}\quad
  \textbf{Yijun Chen\textsuperscript{2}}\quad
  \textbf{Zhiyang Jia\textsuperscript{2}} \\
  \textbf{Jingyu Lu\textsuperscript{1}}\quad
  \textbf{Fan Zhuo\textsuperscript{1}}\quad
  \textbf{Xueyi Pu\textsuperscript{1}}\quad
  \textbf{Yangzhuo Li\textsuperscript{3}}\quad
  \textbf{Zhou Zhao\textsuperscript{1}$^\dagger$} \\
  \vspace{0.4em}
  {\small
  \textsuperscript{1}Zhejiang University \quad
  \textsuperscript{2}China University of Petroleum-Beijing at Karamay \quad
  \textsuperscript{3}Xiamen University
  } \\
  \texttt{leungtianle@gmail.com, zhaozhou@zju.edu.cn} \\
  {\small $^*$ Equal contribution \quad $^\dagger$ Corresponding author}
}

\begin{document}
\maketitle
\begin{abstract}
Recent end-to-end spoken dialogue models enable natural interaction. However, as user demands become increasingly complex, models that rely solely on conversational abilities often struggle to cope. Incorporating agentic capabilities is therefore essential: by enabling tool use, these models can extend their knowledge boundaries and better solve real-world tasks. Yet, existing research has largely concentrated on core perception and generation, with comparatively limited exploration of such tool-augmented extensions. To bridge this gap, we present \textbf{VoxMind}, an integrated framework designed to \textbf{equip end-to-end spoken dialogue models with comprehensive agentic abilities}. Leveraging our curated 470-hour \textbf{AgentChat} dataset, we incorporate a "Think-before-Speak" mechanism, enabling the model to internalize structured reasoning as a critical prerequisite for planning and response generation. Furthermore, to mitigate latency bottlenecks caused by large-scale tool integration, we propose a \textbf{Multi-Agent Dynamic Tool Management architecture}. By asynchronously delegating retrieval tasks to an auxiliary agent aligned with the main model’s reasoning trajectory, this system effectively \textbf{decouples inference latency from toolset size}. Experimental results confirm that VoxMind achieves significant improvements in agent performance: compared with strong baselines, the task completion rate increases from 34.88\% to 74.57\%, outperforming Gemini-2.5-Pro on spoken agent tasks while preserving general conversational quality. The source code and associated data are publicly available at \url{https://github.com/MM-Speech/VoxMind}.
\end{abstract}

\begin{figure}[t]
  \includegraphics[width=\columnwidth]{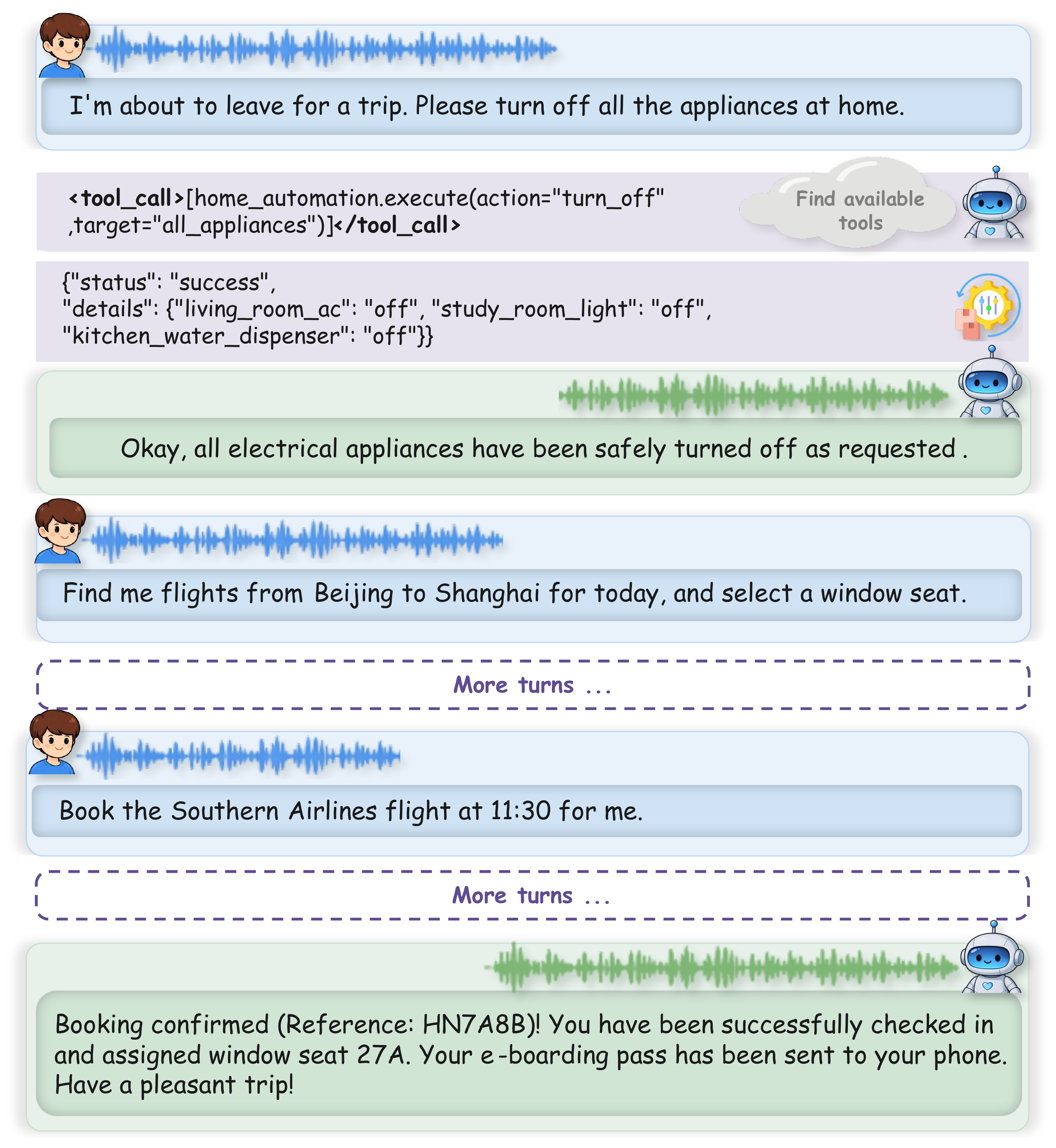}
  \caption{VoxMind can dynamically perceive the interaction context, autonomously determine when to invoke external tools, and drive the generation of subsequent responses based on the tool execution results.}
  \label{fig:first}
\end{figure}

\section{Introduction}

End-to-end spoken dialogue models \cite{Zhang2023SpeechGPTEL,Xie2024MiniOmni2TO,Chen2024SLAMOmniTV,KimiTeam2025KimiAudioTR,Wu2025StepAudio2T,Ji2024WavTokenizerAE,Xu2025Qwen3OmniTR} have emerged as a paradigm shift in speech-based human--computer interaction, as they directly model paralinguistic information and generate expressive spoken responses within the speech modality, instead of using the traditional cascaded ASR‑LLM‑TTS pipeline \cite{ji2024wavchat}. These models have achieved rapid progress in perception and generation, substantially improving naturalness and responsiveness in conversational settings \cite{li2025reinforcement,xu2025qwen2,Long2025VITAAudioFI,Li2025BaichuanAudioAU,chen2026dualaxisgenerativerewardmodel,lu2026modelingbenchmarkingspokendialogue}. Nevertheless, most existing systems remain primarily optimized for reactive conversation\cite{chen-etal-2025-interactspeech,Zhang2024GTSingerAG,chen2026wavalignenhancingintelligenceexpressiveness}, exhibiting limited capacity to handle complex, goal-oriented tasks that require reasoning, planning, and external knowledge access.

Research on text-based agents has shown that mature tool-calling and planning mechanisms can substantially enhance large language models in handling real-time knowledge access and complex reasoning \cite{Schick2023ToolformerLM,Luo2025LargeLM}. In contrast, end-to-end spoken agents remain relatively underexplored and face a set of closely related challenges. At the conceptual level, the speech domain still lacks a unified and widely accepted definition of what constitutes an end-to-end spoken agent, leaving both model design and evaluation without a clear standard. From a capability perspective, end-to-end spoken dialogue models generally lag behind pure text-based models in fine-grained semantic understanding and structured action formulation, such as interpreting tool semantics and generating well-formed tool invocations with appropriate parameters. This limitation directly constrains their ability to support robust planning and long-horizon decision making. The situation is further compounded by the scarcity of speech data explicitly annotated with agentic behaviors, including structured reasoning traces and tool interaction supervision. Moreover, spoken inputs inherently require substantially more tokens to encode rich acoustic information than text, and when combined with large-scale tool descriptions, this results in significant computational overhead, leading to increased inference latency and hindering practical deployment.

To bridge these gaps, we first formulate a rigorous definition of End-to-End Spoken Agents, establishing a unified standard for agentic behaviors in the speech domain. Guided by this formulation, we propose \textbf{VoxMind}, a unified framework that integrates autonomous reasoning, tool utilization, and natural spoken interaction, as illustrated in Fig\ref{fig:first}. To enhance planning capabilities in complex scenarios, VoxMind adopts a \textbf{"Think-before-Speak"} mechanism, enabling the model to perform explicit internal reasoning prior to response generation.

To support reasoning-aware training, we construct the \textbf{AgentChat} dataset, a large-scale spoken corpus annotated with structured reasoning trajectories and tool interaction labels. Training on AgentChat enables VoxMind to internalize cognitive planning processes and generate structured reasoning and tool invocations directly from spoken context.

\begin{figure*}[t]
  \includegraphics[width=0.96\linewidth]{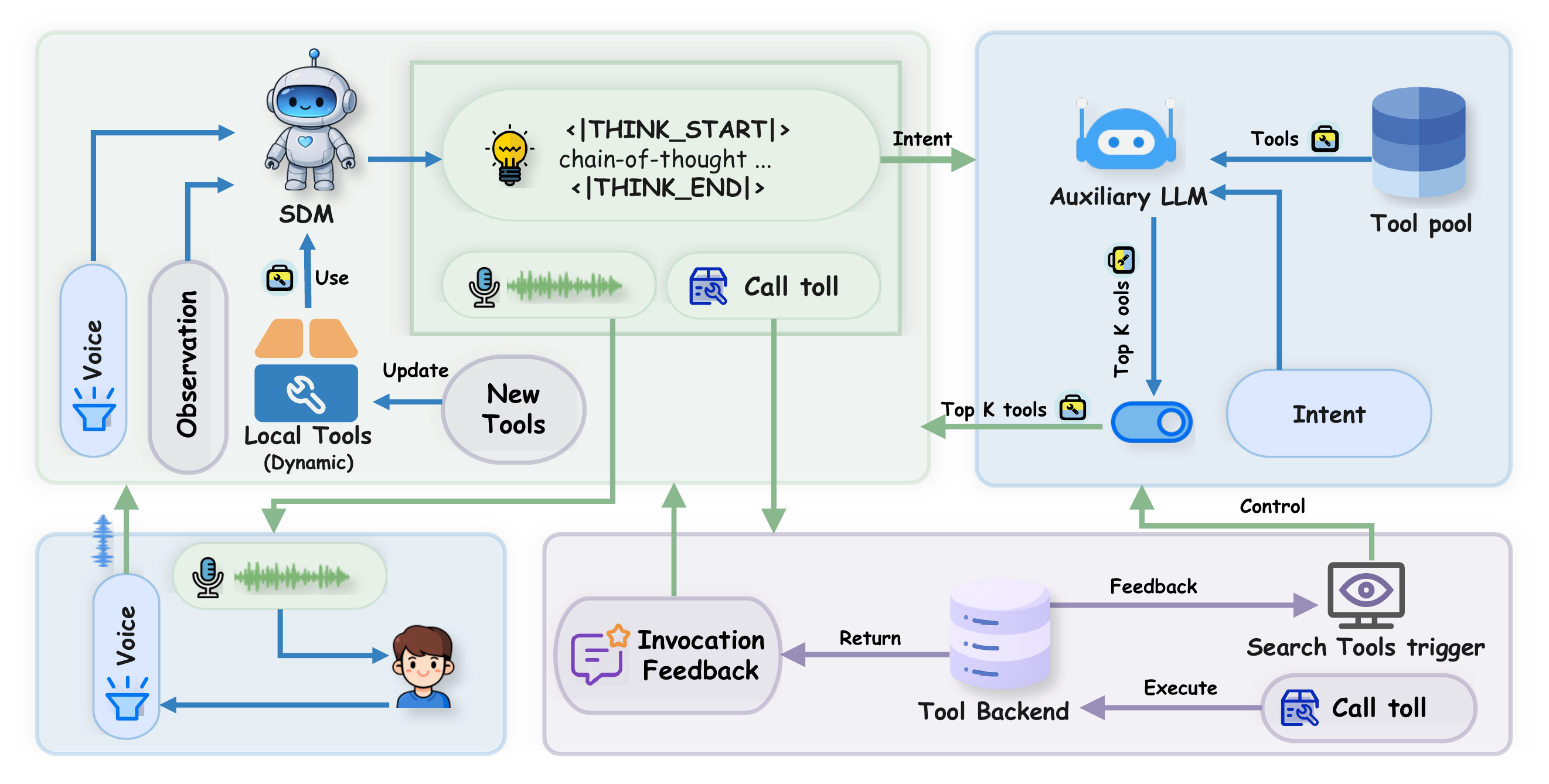} \hfill
  \caption {Overall architecture of the VoxMind. 
Given spoken user input, the speech-centric agent first generates an explicit reasoning trajectory in a "think-before-speak" manner. Conditioned on this reasoning output, the speech model generates a response, while an auxiliary language model operates in parallel to propose candidate tools from a global pool. The selected action and the proposed tool set jointly determine dynamic updates to the agent’s local tool space, enabling scalable tool usage without increasing response latency.}
  \label{fig:second} 
\end{figure*}

In addition, to enable scalable tool usage with low latency, VoxMind incorporates a \textbf{Dynamic Tool Management} mechanism based on a multi-agent design. The system maintains a compact, reasoning-conditioned local tool space that is dynamically updated with candidate tools selected from a global pool, thereby avoiding repeated processing of the entire tool library. This design effectively decouples inference efficiency from toolset scale, enabling responsive decision making in tool-rich environments.

In summary, our main contributions are as follows:
  \begin{itemize}
    \item We formulate a formal definition for End-to-End Spoken Agents, bridging a critical theoretical gap in the field. Building on this foundation, we propose \textbf{VoxMind}, a unified model that incorporates a "Think-before-Speak" paradigm to effectively execute these complex reasoning and tool-use tasks.
    \item We construct \textbf{AgentChat}, a speech dataset explicitly annotated with reasoning trajectories, tool interactions, and complex planning paths. This resource alleviates the scarcity of agentic supervision in spoken contexts, facilitating the development of reasoning-aware speech agent.
    \item We design a \textbf{Multi-Agent Dynamic Tool Management architecture} that employs an asynchronous parallel execution strategy. This mechanism decouples inference latency from the size of the tool library, ensuring consistent performance and accuracy as the toolset expands.
  \end{itemize}

\section{Related Work}
The reliance of pre-trained large language models (LLMs) on static training data limits their adaptability to dynamic scenarios~\cite{Qu2024ToolLW}. The autonomous agent paradigm mitigates this by enabling models to interface with external tools~\cite{Masterman2024TheLO}. While reasoning frameworks are well-established in the text domain~\cite{Yao2022ReActSR,Qin2023ToolLLMFL,Hong2023MetaGPTMP}, the extension to end-to-end voice interaction remains nascent. Recent works, including Stream RAG~\cite{Arora2025StreamRI},WavRAG~\cite{Chen2025WavRAGAR}, TARL~\cite{Tan2025ProcessSupervisedRL}, and Qwen3-Omni~\cite{Xu2025Qwen3OmniTR}, demonstrate preliminary agent capabilities. However, these efforts lack systematic exploration, primarily limiting models to isolated functionalities such as information retrieval or basic tool use. Consequently, solving complex problems necessitates a comprehensive system architecture, as simple functional extensions are insufficient.

\section{Methodology}

\subsection{Unified Definition of End-to-End Spoken Agents}

We define an \textbf{End-to-End Spoken Agent} as an autonomous system that transcends reactive speech generation to possess cognitive and executable capabilities. To facilitate complex problem-solving in spoken scenarios, we formulate the agent $\mathcal{A}$ as a unified framework consisting of four essential dimensions.

\textbf{Profile Definition.} A comprehensive agent profile must encompass both semantic roles and acoustic identities. We decompose this definition $\mathcal{P}$ into two dimensions to balance consistency with adaptability: \textbf{Static Definition (Consistency):}  This specifies the agent's inherent attributes, denoted as $P_{static}$, such as timbre, gender, age, accent, and semantic persona (e.g., customer service agent, educational expert). These features are pre-defined to maintain a cohesive persona throughout interactions, ensuring the user perceives a stable conversational partner. \textbf{Dynamic Adaptive Definition (Autonomy):} This encompasses the agent's self-definition $P_{dynamic}$, derived from real-time environmental interaction and self-reflection. Attributes such as emotional tone, speaking rate, rhythm, and prosody are not hard-coded; rather, they are dynamically determined by the agent in response to the context $c$ (e.g., sensing user urgency). This mechanism reflects the agent's situational awareness and autonomy, formalized as $\mathcal{P} = ( P_{static}, P_{dynamic}(c) )$.

\textbf{Memory Mechanism.} To overcome the base model's inherent statelessness, a robust memory mechanism needs to be introduced to persist interactions across time.This mechanism enforces a \textbf{dual-channel architecture} throughout all storage levels, maintaining both \textbf{Semantic Memory} ($\mathcal{M}_{sem}$) and \textbf{Acoustic Memory} ($\mathcal{M}_{acous}$) to capture what was said and how it was said.\textbf{Short-Term Memory ($\mathcal{M}_{ST}$):} Functioning as working memory, this module buffers the immediate multi-modal context. It simultaneously retains semantic content and paralinguistic acoustic features (e.g., emotion, pitch), enabling the agent to maintain situational awareness in real-time fluid interactions.\textbf{Long-Term Memory:} This component archives persistent knowledge accumulated over extended periods. It stores not only historical facts and user preferences (Semantic) but also recurring vocal patterns and prosodic habits (Acoustic), ensuring long-term consistency in the agent's interaction style. 

\textbf{Planning Capability.} To solve complex real-world problems, the agent cannot rely solely on reactive behavior (i.e., reflexive response). While end-to-end models typically perform a direct mapping from input to output ($x \to y$), this formulation is often insufficient for complex planning tasks. Thus, an effective agent requires an intermediate reasoning stage $z$, transforming the interaction paradigm into $x \to z \to y$. Here, $x \in \mathcal{X}$ denotes the multimodal input, $z \in \mathcal{Z}$ represents the intermediate reasoning process (e.g., chain-of-thought, task decomposition, or latent logic generation), and $y \in \mathcal{Y}$ corresponds to the final executed action or spoken response. This intermediate step $z$ enables the agent to deliberate and formulate a structured plan prior to execution.

\textbf{Action Execution.} Planning alone remains theoretical without execution. Therefore, this principle centers on \textbf{Tool Utilization}, where the execution process is governed by two sequential decision-making stages. \textbf{Decision:} The agent evaluates the current context to determine if external assistance is necessary to fulfill the plan. \textbf{Selection And Invocation:} Upon confirming the need for tools, the agent identifies the optimal tool $t^*$ from the available API set $\mathcal{T}$ and generates the precise parameters required for invocation.

\subsection{VoxMind}

To construct a comprehensive spoken dialogue agent, we propose the VoxMind architecture, as shown in Fig\ref{fig:second}. The system state at time step $t$ is defined as:
\begin{equation}
  \label{eq:state}
  \mathcal{S}_t = (\mathbf{O}_t, \mathcal{H}_t, \mathcal{A}_t)
\end{equation}
where $\mathbf{O}_t$ denotes the set of observable events at time $t$, comprising the current user input $\mathbf{X}_t$ and structured feedback $\mathbf{O}_t^{env}$ returned by tools or the environment (i.e., $\mathbf{O}_t = \{\mathbf{X}_t, \mathbf{O}_t^{env}\}$). $\mathcal{H}_t$ represents the accumulated interaction history, and $\mathcal{A}_t$ denotes the agent's action space, consisting of verbal responses $\mathcal{V}$ and a dynamically retrieved subset of callable tools $\mathcal{T}_t^{local} \subset \mathcal{T}^{all}$.

The core objective of VoxMind is to learn a hierarchical policy that maps the system state $\mathcal{S}_t$ to an optimal action $\mathbf{a}_t \in \mathcal{A}_t$ via an explicit "think-before-speak" mechanism. Specifically, before producing speech output or invoking tools, the agent generates an explicit \textbf{Chain-of-Thought (CoT) reasoning trajectory}:
\begin{equation}
\mathbf{c}_t \sim \pi_{\theta}^{\text{think}}\!\left(\mathbf{c} \mid \mathbf{o}_t,\ \mathcal{H}_{t-1},\ \mathcal{T}_t^{local}\right).
\end{equation}
This trajectory captures intent understanding, contextual analysis, and task planning, ensuring that the reasoning step is completed prior to any action execution.

Conditioned on the sampled reasoning trajectory, the agent selects its next action based on the current observation, interaction history, and locally accessible tools:
\begin{equation}
\mathbf{a}_t \sim \pi_{\theta}^{\text{act}}\!\left(\mathbf{a} \mid \mathbf{c}_t,\ \mathbf{o}_t,\ \mathcal{H}_{t-1},\ \mathcal{T}_t^{local}\right).
\end{equation}
The resulting action corresponds either to a verbal response or the invocation of an external tool. This ensures that all observable behaviors are grounded in explicit reasoning while remaining consistent with the current context and tool availability.

To effectively decouple inference latency from toolset size for scalable tool usage, VoxMind employs a \textbf{parallel dynamic tool update mechanism} driven by an auxiliary language model. After the reasoning trajectory $\mathbf{c}_t$ is generated, the system executes two processes in parallel: the agent samples its next action conditioned on the current local tool set, while an auxiliary model proposes candidate tools from the global pool:
\begin{equation}
\resizebox{0.9\linewidth}{!}{$
    (\mathbf{a}_t,\ \mathcal{T}_t^{cand})
    \sim
    \big(\pi_{\theta}^{\text{act}}(\cdot \mid \mathbf{c}_t,\mathcal{T}_t^{local}),\ 
    \pi_{\text{LLM}}(\mathbf{c}_t,\ \mathcal{T}^{all})\big)
$}
\end{equation}

The sampled action explicitly determines the state transition of the tool space. When the agent emits the retrieval action $\mathbf{a}_t = a_{\text{retrieve}}$, indicating that the current local tool set is insufficient to accomplish the task, the candidate tools proposed by the auxiliary model are incorporated to form the tool space for the next decision step:
\begin{equation}
\mathcal{T}_{t+1}^{local}
=
\mathcal{T}_t^{local} \cup \mathcal{T}_t^{cand}.
\end{equation}
Otherwise, the local tool set remains unchanged, i.e., $\mathcal{T}_{t+1}^{local} = \mathcal{T}_t^{local}$.

Conditioned on the updated tool availability, the agent then performs its next decision at time step $t{+}1$ to obtain the final executable action:
\begin{equation}
\mathbf{a}_{t+1}
\sim
\pi_{\theta}^{\text{act}}\!\left(\mathbf{a} \mid \mathbf{c}_{t+1},\ \mathbf{o}_{t+1},\ \mathcal{H}_{t},\ \mathcal{T}_{t+1}^{local}\right).
\end{equation}
This design enables scalable tool usage by explicitly triggering tool expansion upon detected insufficiency with minimal inference overhead.

\subsection{AgentChat}

\textbf{Basic Interaction Data Construction.} To train a robust intelligent agent, we constructed the AgentChat dataset, which comprises two distinct corpora (as shown in Table \ref{tab:AgentChat}): a \textbf{tool‑interaction corpus} and a \textbf{general‑conversation corpus}. The construction process involves rigorous text collection, cleaning, and speech synthesis; each stage is detailed below.

The tool‑interaction corpus is derived from existing benchmark datasets, including ToolACE~\cite{Liu2024ToolACEWT} and APIGen‑MT~\cite{Prabhakar2025APIGenMTAP}. We first perform coarse rule‑based filtering to remove content unsuitable for speech synthesis, such as HTML tags, Markdown markers, and code snippets. Subsequently, fine‑grained filtering is carried out using the Qwen‑plus\footnote{https://bailian.console.aliyun.com} language model, polishing the text to ensure a natural conversational style while removing data inappropriate for speech scenarios. To further enrich the tool data, we also employed the language model to generate a set of task‑specific dialogues based on the established tool descriptions from ToolACE~\cite{Liu2024ToolACEWT}.

The general‑conversation corpus integrates publicly available datasets (SciQ~\cite{Welbl2017CrowdsourcingMC}, GSM8K~\cite{Cobbe2021TrainingVT}, ARC~\cite{Clark2018ThinkYH}) as well as data derived from common knowledge found in secondary school textbooks. We selected subsets suitable for speech synthesis, resulting in a domain‑balanced collection.

All cleaned text is converted to speech using CosyVoice2. To increase speaker diversity and acoustic naturalness, we utilized over 600 prompt‑based timbres from SeedTTS~\cite{Anastassiou2024SeedTTSAF} during synthesis, producing a stylistically diverse and high‑fidelity speech corpus.

\begin{table}
  \centering
  \resizebox{\columnwidth}{!}{%
  \definecolor{TableGroupGray}{RGB}{240, 240, 240}
  \begin{tabular}{lrrr}
    \hline
    \textbf{AgentChat} & \textbf{Instances} & \textbf{Avg. Turns} & \textbf{Duration (H)} \\
    \hline
    \rowcolor{TableGroupGray}
    \multicolumn{4}{l}{\textbf{Tool Interaction Data}} \\
    tool-ace-audio         & 5582  & 1.0672 & 26.6220 \\
    apigen-mt-audio        & 791   & 7.4355 & 43.2587 \\
    Self-built (Tool)      & 8432  & 1.3455 & 39.1885 \\
    \hline
    \rowcolor{TableGroupGray}
    \multicolumn{4}{l}{\textbf{General Dialogue Data}} \\
    ai2\_arc-challenge     & 1167  & 1.0000 & 12.3334 \\
    ai2\_arc-easy          & 1164  & 1.0000 & 10.8154 \\
    gsm8k                  & 1746  & 1.0000 & 18.4730 \\
    sciq                   & 998   & 1.0000 & 9.4903  \\
    Self-built (Normal)    & 26406 & 1.1201 & 309.8364 \\
    \hline
  \end{tabular}%
  }
  \caption{\label{tab:AgentChat}
    Composition of the AgentChat dataset. AgentChat consists of \textbf{Tool Interaction Data} (comprising the \textit{tool-ace-audio}, \textit{apigen-mt-audio}, and \textit{Self-built (Tool)} subsets) and \textbf{General Dialogue Data} (comprising the \textit{ai2\_arc-challenge}, \textit{ai2\_arc-easy}, \textit{gsm8k}, \textit{sciq}, and \textit{Self-built (Normal)} subsets).
  }
\end{table}

\noindent
\textbf{Chain-of-Thought Construction.} To construct intermediate reasoning trajectories for training, we adopt a reverse conditional generation approach. Specifically, given a task input $Q$ and the corresponding final output $A$, the model generates a reasoning chain $R$ that logically bridges them. This process is formulated as sampling from the conditional distribution:
\begin{equation}
R \sim p_{\text{LM}}(R \mid Q, A).
\end{equation}

To ensure quality, we implement an iterative filtering mechanism based on scoring. Each candidate reasoning chain $R$ is assigned a quality score $S(R) \in [0, 10]$. Only chains satisfying a predefined threshold $\tau = 7$ are retained:
\begin{equation}
\mathcal{R}_{\text{retain}} = \{ R \mid S(R) \geq \tau \}.
\end{equation}
For chains falling below the threshold, the system regenerates the reasoning chain up to $T=3$ times:
\begin{equation}
\begin{array}{c}
R_{i+1} \sim p_{\text{LM}}(R_i \mid Q, A), \\
i \in \{i' \mid i' \leq 2, S(R_{i'}) < \tau\}.
\end{array}
\end{equation}
Candidates that fail to meet the threshold after three attempts are discarded.

Finally, each retained chain undergoes textual refinement. A large language model polishes the reasoning text to improve conciseness and standardize the format. Guided by instructions $\mathcal{I}$, this process strictly preserves the core logical flow:
\begin{equation}
R' = \text{LLM}_{\text{refine}}(R \mid \mathcal{I}).
\end{equation}
The resulting dataset of refined chains, ${R'}$, provides clean and structured trajectory data for effective agent training.

\begin{figure*}[t]
  \centering
  \scalebox{1}[0.93]{\includegraphics[width=0.96\linewidth]{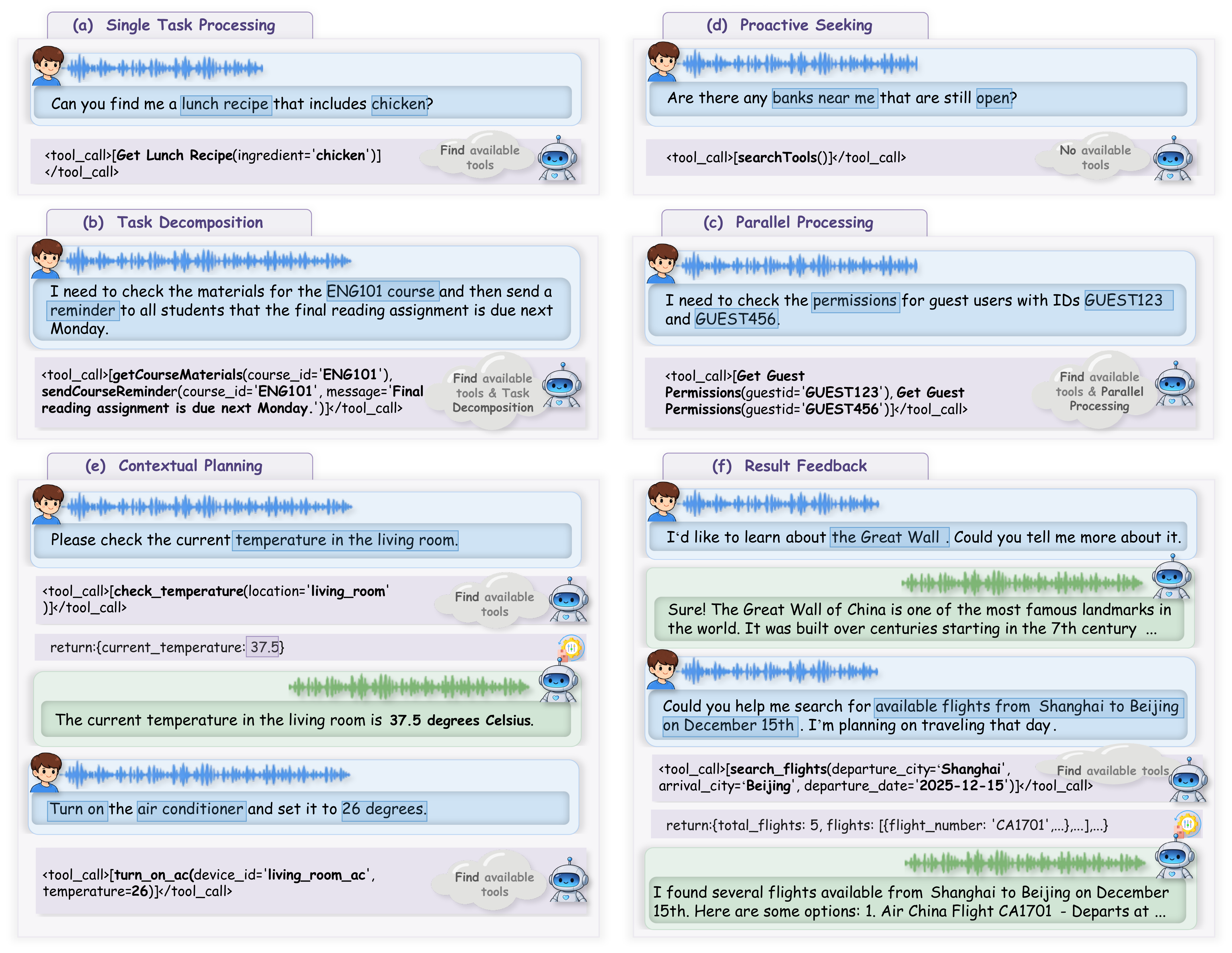}} \hfill
  
  \caption {Dialogues demonstrating the agent's six core capabilities.}
  \label{fig:third} 
\end{figure*}

\begin{table*}[t]
\centering
\resizebox{\textwidth}{!}{%
\begin{tabular}{l
cc cc cc cc
c c c}
\toprule

\textbf{Model}
& \multicolumn{2}{c}{\textbf{Single Task Processing}}
& \multicolumn{2}{c}{\textbf{Task Decomposition}}
& \multicolumn{2}{c}{\textbf{Parallel Processing}}
& \multicolumn{2}{c}{\textbf{Contextual Planning}}
& \textbf{Proactive Seeking}
& \textbf{Result Feedback}
& \textbf{Overall} \\

\cmidrule(lr){2-3}
\cmidrule(lr){4-5}
\cmidrule(lr){6-7}
\cmidrule(lr){8-9}
\cmidrule(lr){10-10}
\cmidrule(lr){11-11}
\cmidrule(lr){12-12}

& TS$\uparrow$ & PF$\uparrow$
& TS$\uparrow$ & PF$\uparrow$
& TS$\uparrow$ & PF$\uparrow$
& TS$\uparrow$ & PF$\uparrow$
& TU$\uparrow$
& FC$\uparrow$
&  \\

\midrule

\rowcolor{gray!12}
\multicolumn{12}{l}{\textbf{Closed-source models (direct inference with prompt)}} \\

Gemini-2.5-pro
& 90.98 & \underline{\underline{75.19}}
& \underline{\underline{82.54}} & \textbf{52.38}
& \underline{\underline{88.57}} & \textbf{69.52}
& \underline{\underline{84.25}} & \underline{61.64}
& \underline{26.87}
& \underline{\underline{4.16}}
& \underline{\underline{71.51}} \\

Gemini-2.5-flash
& \underline{92.48} & \textbf{77.44}
& \underline{61.90} & 31.22
& \underline{86.67} & \underline{\underline{68.25}}
& \textbf{86.99} & \textbf{65.75}
& \underline{\underline{31.34}}
& \underline{4.10}
& \underline{68.40} \\

GPT-4o-audio
& 85.71 & 70.68
& 23.81 & 15.87
& 84.76 & \underline{61.90}
& 71.23 & 49.32
& 0.00
& \textbf{4.22}
& 54.77 \\

\midrule

\rowcolor{gray!12}
\multicolumn{12}{l}{\textbf{Open-source models (direct inference with prompt)}} \\

\rowcolor{gray!8}
\multicolumn{12}{l}{\quad \textit{Cascaded models}} \\

Qwen3-8B+Whisper
& \underline{\underline{94.99}} & 68.42
& \underline{\underline{82.54}} & \underline{\underline{41.27}}
& 85.71 & 46.67
& \underline{\underline{84.25}} & 47.72
& 7.46
& 4.05
& 64.00 \\

\rowcolor{gray!8}
\multicolumn{12}{l}{\quad \textit{End-to-end models}} \\

Kimi-Audio
& 78.45 & 56.89
& 48.15 & 22.75
& 79.05 & 55.24
& 76.03 & 46.80
& 13.64
& 3.62
& 54.94 \\

Qwen2.5-Omni
& 78.70 & 35.84
& 38.62 & 3.17
& 65.40 & 28.57
& 65.75 & 26.03
& 0.00
& 2.82
& 39.85 \\

StepAudio2
& 78.70 & 48.87
& 60.32 & 26.98
& 53.33 & 33.33
& 4.34 & 1.60
& 3.12
& 1.91
& 34.88 \\

\midrule

\rowcolor{gray!12}
\multicolumn{12}{l}{\textbf{Ours}} \\

\textbf{VoxMind}
& \textbf{98.50} & \underline{72.18}
& \textbf{95.24} & \underline{38.10}
& \textbf{89.52} & 61.59
& \underline{80.82} & \underline{\underline{62.33}}
& \textbf{68.66}
& 3.94
& \textbf{74.57} \\

\bottomrule
\end{tabular}
}
\caption{We evaluate model performance using four metrics: TS (Tool Selection accuracy), PF (Parameter Filling accuracy), TU (Tool Usage accuracy), and FC (Feedback Completeness).}
\label{tab:agent_speech_results}
\end{table*}

\begin{table}[t]
\centering
\resizebox{\columnwidth}{!}{%
\begin{tabular}{ll cccc}
\toprule

\multicolumn{2}{l}{\multirow{2}{*}{\textbf{Metric}}} 
& \textbf{w/o think} & \textbf{w/o think} & \textbf{w/ think} & \textbf{w/ think} \\
\multicolumn{2}{c}{} 
& \textbf{(1:1)} & \textbf{(1:0.5)} & \textbf{(1:1)} & \textbf{(1:0.5)} \\

\midrule

\multirow{2}{*}{\textbf{Single Task Processing}} 
& TS$\uparrow$ 
& 88.72 & 90.23 & \underline{90.98} & \textbf{98.50} \\
& PF$\uparrow$ 
& 70.68 & \underline{71.68} & 68.42 & \textbf{72.18} \\

\addlinespace

\multirow{2}{*}{\textbf{Task Decomposition}} 
& TS$\uparrow$ 
& \textbf{95.24} & 93.65 & \underline{94.71} & \textbf{95.24} \\
& PF$\uparrow$ 
& \underline{39.68} & 36.51 & \textbf{44.44} & 38.10 \\

\addlinespace

\multirow{2}{*}{\textbf{Parallel Processing}} 
& TS$\uparrow$ 
& 80.00 & 80.00 & \underline{80.95} & \textbf{89.52} \\
& PF$\uparrow$ 
& 45.71 & \underline{59.05} & 51.43 & \textbf{61.59} \\

\addlinespace

\multirow{2}{*}{\textbf{Contextual Planning}} 
& TS$\uparrow$ 
& \textbf{86.99} & \underline{86.30} & 84.93 & 80.82 \\
& PF$\uparrow$ 
& \underline{73.29} & \textbf{75.34} & 65.75 & 62.33 \\

\addlinespace

\textbf{Proactive Seeking} 
& TU$\uparrow$ 
& 31.34 & 37.31 & \underline{59.70} & \textbf{68.66} \\

\textbf{Result Feedback} 
& FC$\uparrow$ 
& 3.83 & \textbf{3.98} & 3.92 & \underline{3.94} \\

\addlinespace

\textbf{Overall} 
& 
& 68.83 & 70.97 & \underline{71.97} & \textbf{74.57} \\

\bottomrule
\end{tabular}
}
\caption{Ablation Study. Investigate the impact of deep reasoning on agent performance. The metrics are preserved across different training strategies.}
\label{tab:ablation}
\end{table}

\noindent
\textbf{Core Agent Competencies.} Our method equips the agent with a suite of core capabilities through targeted training on the AgentChat dataset, as illustrated in Fig\ref{fig:third}. The design encompasses the following key functions: \textbf{single-task processing}, enabling the agent to accurately understand user intent and invoke appropriate tools for independent tasks; \textbf{task decomposition}, allowing it to break down complex requests into manageable subtasks; \textbf{parallel processing}, which enhances efficiency by identifying independent subtasks of the same type and generating parallel execution plans; \textbf{proactive seeking}, empowering the agent to initiate external searches or requests when existing tools are inadequate, thus adapting to open-world scenarios; \textbf{result feedback}, which enables dynamic adjustment of subsequent actions based on tool execution outcomes; \textbf{contextual planning}, leveraging historical interaction context to maintain coherence in multi-turn dialogues.  See Appendix \ref{sec:Detailed_Composition_of_the_Dataset} for details on the dataset composition.

\section{Experiments}

\subsection{Experimental Setup}

\noindent
\textbf{Datasets.} Given the lack of open-source agent interaction data in speech environments, we train our model on the AgentChat dataset. We reserve a \textbf{disjoint subset} as an independent test set to evaluate the agent's core capabilities. Additionally, we construct an \textbf{out-of-domain} dataset using Gemini-2.5-Pro to investigate model performance across \textbf{expanding tool scales} in real-world scenarios. To explore the impact of data proportioning on agent training effectiveness, we further generate two datasets with \textbf{distinct ratio configurations} (1:1 and 1:0.5), where the ratio denotes the time proportion between speech-oriented agent interaction data and general dialogue data. Complete dataset statistics and composition details are provided in Appendix \ref{sec:train_config}.

\noindent
\textbf{Baselines.} For extensive comparison, we select a suite of competitive models, including leading closed-source models like Gemini-2.5-pro\footnote{https://ai.google.dev/gemini-api}, Gemini-2.5-flash, and GPT-4o-audio\footnote{https://platform.openai.com/docs/models/gpt-4o-audiopreview}, as well as open-source ones such as Qwen2.5-Omni\cite{Xu2025Qwen25OmniTR}, Kimi-Audio\cite{KimiTeam2025KimiAudioTR}, and Qwen3+Whisper\cite{Yang2025Qwen3TR, Radford2022RobustSR}. StepAudio2\cite{Wu2025StepAudio2T}, also an open-source model, serves as the foundation for fine-tuning.

\noindent
\textbf{Evaluation Setup.}
Our evaluation covers three complementary aspects. We first assess six core agent capabilities illustrated in Fig.\ref{fig:third}: single-task processing, task decomposition, parallel processing, proactive seeking, result feedback, and contextual planning. These capabilities are quantified using four task-level metrics: \textbf{TS} (Tool Selection accuracy), evaluating correct tool selection from the local tool set; \textbf{PF} (Parameter Filling accuracy), measuring structured parameter instantiation from context; \textbf{TU} (Tool Usage accuracy), assessing the agent’s ability to detect tool insufficiency and trigger retrieval; and \textbf{FC} (Feedback Completeness), evaluating accurate perception and summarization of environment feedback. All evaluations are conducted using \textbf{Gemini-2.5-Flash} as an expert evaluator by \textbf{verifying model outputs against predefined ground-truth answers}, rather than subjective scoring. To improve robustness and reduce evaluator variance, \textbf{each model output is evaluated three times and the final score is obtained by averaging}, with detailed evaluation prompts and criteria provided in Appendix \ref{sec:Evaluation_Core_Competencies}.

We additionally evaluate \textbf{VoxMind} on the \textbf{VoiceBench}~\cite{Chen2024VoiceBenchBL} benchmark to verify that general conversational ability is preserved under agentic training.

Finally, to analyze the impact of dynamic tool management, we conduct controlled experiments on a \textbf{Gemini-generated cross-domain dataset}, comparing configurations \textbf{with and without the auxiliary tool management agent} while varying the number of available tools and measuring \textbf{task accuracy and relative inference latency}.

\noindent
\textbf{Training details.} The experiments were configured with the following key parameters: we adopted 2 pieces of H20-NVLink GPU for model training . The batch size was set to 1, with gradient accumulation steps of 8 to compensate for the small batch size. The learning rate was initialized at 1e-5, and a cosine learning rate scheduler was employed during training. Other regularization and optimization settings included a weight decay of 0.01, a maximum gradient norm clipping of 1.0, and the AdamW optimizer. For efficient large-scale model training, we enabled DeepSpeed with the ZeRO-3 strategy, bfloat16 precision, and gradient checkpointing.Further implementation details can be found in Appendix \ref{sec:train_config} .

\subsection{Results and Analysis}

\begin{figure*}[t]
  \centering
  \includegraphics[width=0.96\linewidth]{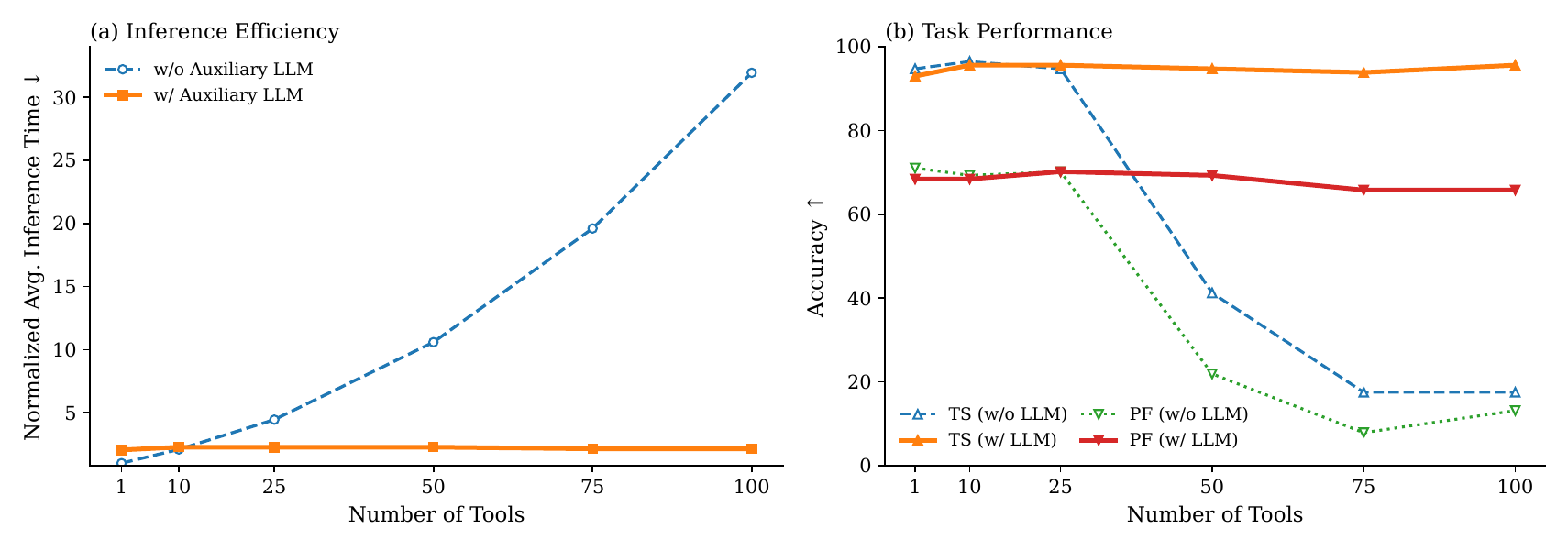}
  \caption{
Comparison of inference efficiency and task accuracy with and without the auxiliary LLM across varying tool pool sizes. The auxiliary LLM enables efficient tool-space pruning, significantly reducing inference overhead while maintaining performance.
  }
  \label{fig:muti-agent}
\end{figure*}

\noindent
\textbf{Evaluation of Core Capabilities of Agents.} As shown in Table \ref{tab:agent_speech_results}, VoxMind achieves SOTA performance with an overall score of 74.57. Compared to the base model StepAudio2 (34.88), VoxMind demonstrates a substantial relative improvement of 113.79\%. It also significantly outperforms the top-tier open-source end-to-end model, Kimi-Audio (54.94), and the cascaded system Qwen3-8B + Whisper (64.00), while surpassing even the leading closed-source model, Gemini-2.5-pro (71.51).

Among open-source baselines, the cascaded system (Qwen3-8B + Whisper) notably outperforms end-to-end alternatives such as Kimi-Audio. This suggests that cascaded systems leveraging text-based LLMs maintain an advantage when paralinguistic information and latency are not primary constraints. Additionally, a broader comparison reveals that closed-source models generally outperform open-source baselines, highlighting a discernible gap between community-driven and proprietary models.

Collectively, these results indicate that current end-to-end speech large models still exhibit suboptimal performance on agentic tasks. This highlights the importance of our proposed VoxMind, which serves as a vital contribution to the open-source community.

\begin{table}[t]
\centering
\resizebox{\columnwidth}{!}{%
\begin{tabular}{lccccc}
\toprule
\textbf{VoiceBench}
& \makecell{\textbf{Step-Audio-2}\\\textbf{(Base)}}
& \makecell{\textbf{w/o think}\\\textbf{(1:0.5)}}
& \makecell{\textbf{w/o think}\\\textbf{(1:1)}}
& \makecell{\textbf{w/ think}\\\textbf{(1:1)}}
& \makecell{\textbf{w/ think}\\\textbf{(1:0.5)}} \\

\midrule

AlpacaEval        & \textbf{4.19} & 3.38 & 3.77 & \underline{4.08} & 3.98 \\
CommonEval        & 3.12 & 3.43 & 3.75 & \textbf{4.03} & \underline{3.94} \\
WildVoice         & 3.36 & 3.02 & 3.42 & \textbf{3.79} & \underline{3.69} \\

\midrule
SD-QA (USA) / Panda
                  & \textbf{55.15} & 49.73 & 48.28 & \underline{51.90} & 49.73 \\
SD-QA (USA) / GPT
                  & \textbf{52.80} & 38.34 & 39.24 & 44.48 & \underline{44.85} \\

\midrule
MMSU               & 50.82 & 36.88 & 47.69 & \underline{51.61} & \textbf{53.04} \\
OBQA               & 68.13 & 56.70 & \underline{68.79} & 65.49 & \textbf{71.87} \\
BBH                & \textbf{58.53} & 50.66 & 50.25 & \underline{56.31} & 54.69 \\
IFEval             & \textbf{39.64} & 20.74 & \underline{23.61} & 17.40 & 18.83 \\
AdvBench           & 92.88 & 87.69 & 84.62 & \underline{95.58} & \textbf{100.00} \\

\midrule
\textbf{Overall}
                   & \underline{64.15} & 54.80 & 59.72 & 63.62 & \textbf{64.21} \\
\bottomrule
\end{tabular}
}

\caption{
Performance comparison on the \textbf{VoiceBench} general conversation task between the base model, models with and without deep thinking training, and models trained with different data ratios.
}
\label{tab:voicebench}
\end{table}

\noindent
\textbf{Ablation Study.} Table~\ref{tab:ablation} analyzes the impact of training strategies and data ratios on agent task performance. Without chain-of-thought reasoning (w/o think), increasing the proportion of agent interaction data (shifting from 1:1 to 1:0.5) yields only marginal gains, improving the score from 68.83 to 70.97. This indicates that the direct speech-to-answer paradigm faces a performance bottleneck. Furthermore, Table~\ref{tab:voicebench} reveals that this approach comes at the cost of general speech capability, with the VoiceBench score regressing significantly from 59.72 to 54.80.

Conversely, models incorporating deep reasoning (w/ think) demonstrate superior robustness and performance. The w/ think (1:0.5) configuration achieves a peak agent task score of 74.57 (+2.6 points over the 1:1 baseline) and elevates the general evaluation to 64.21, outperforming both the 1:1 variant and the base model (64.15). Notably, while w/o think models suffer substantial regressions on VoiceBench (dropping between 4.43 and 9.35 points), the w/ think variants exhibit negligible degradation (maximum 0.53 points). This confirms that reasoning capabilities effectively preserve general knowledge while enhancing specialized performance.

These findings suggest that explicit chain-of-thought reasoning is critical for stabilizing training and mitigating the trade-off between domain specialization and general speech proficiency.

\noindent
\textbf{Dynamic Tool Management Analysis.} As shown in Fig \ref{fig:muti-agent}(a), when configured with a single tool, VoxMind's inference time is marginally higher than that of the single-agent approach. However, as the number of tools increases, the single-agent's inference time exhibits exponential growth, rendering it entirely unsuitable for real-world scenarios involving numerous tools. In contrast, VoxMind maintains stable inference times through its auxiliary agent-based tool management mechanism, achieving true decoupling between tool quantity and inference duration. Experimental results in Fig \ref{fig:muti-agent}(b) further validate this advantage: as tool scale expands, single-agent inference performance degrades significantly, whereas the VoxMind model consistently sustains stable performance.

Beyond the above analyses, we conduct additional experiments to further validate our design from three perspectives: robustness to real-world speech, latency-scale decoupling, and token-level overhead. Overall, the results consistently show that our system maintains strong robustness under realistic conditions while introducing only minimal and bounded overhead. A brief summary is provided here, with full experimental details deferred to Appendix~\ref{sec:appendix_real_speech}, Appendix~\ref{sec:appendix_latency}, and Appendix~\ref{sec:appendix_token}, respectively.

\section{Conclusion}

In this work, we establish a comprehensive definition and theoretical standard for End-to-End Spoken Agents. Building on this foundation, we propose VoxMind, an end-to-end spoken agent capable of intrinsic reasoning and tool use. Experimental results demonstrate that VoxMind significantly outperforms strong baselines on complex agentic tasks, providing a robust theoretical and technical framework for the field.

\section*{Limitations}

Despite the advancements presented, two aspects warrant further discussion. First, the core "Think-before-Speak" mechanism, while pivotal for enabling complex reasoning, inherently introduces an inference latency trade-off. The generation of internal reasoning trajectories precedes the final verbal response, inevitably incurring a computational overhead compared to shallow reactive models. We regard this as a necessary trade-off for correctness, yet minimizing this latency remains an objective for future research.Second, regarding dataset construction, the AgentChat dataset relies on synthesizing mature text-based reasoning corpora. Although we implemented rigorous filtering to ensure audio-text alignment, the semantic structure may still reflect the precision of written language rather than the spontaneity and disfluencies characteristic of authentic daily speech. Future iterations will focus on constructing datasets natively rooted in spoken scenarios to better capture the nuances of acoustic pragmatics.

\section{Acknowledgements}

This work was supported by National Natural Science Foundation of China under Grant No.U25B2064


\bibliography{custom}

\appendix

\section{Detailed Composition of the Dataset} 

\label{sec:Detailed_Composition_of_the_Dataset}

\subsection{AgentChat Data Details} 

\begin{table*}[t]
  \centering
  \setlength{\tabcolsep}{6pt}
  \begin{tabular}{l r r r r}
    \toprule
    \textbf{AgentChat-Tool} 
    & \textbf{Samples} 
    & \textbf{Tool Categories} 
    & \textbf{Avg. Turns} 
    & \textbf{Duration (H)} \\
    \midrule
    tool-select & 1,237 & 5,038 & 1 & 1.9225 \\
    multi-tool-select & 1,486 & 7,240 & 1 & 5.1605 \\
    para-filled & 1,409 & 3,626 & 1 & 4.4508 \\
    parallel-call & 1,144 & 4,767 & 1 & 2.5235 \\
    searchTool & 467 & 2,164 & 1 & 0.6172 \\
    tool-ace-audio & 5,582 & 10,892 & 1.0672 & 26.6220 \\
    apigen-mt & 791 & 3,980 & 7.4355 & 43.2587 \\
    observation & 2,465 & 8,423 & 2 & 22.4615 \\
    obs\_searchtools & 224 & 855 & 3 & 2.0525 \\
    \midrule
    \textbf{All} & \textbf{14,805} & \textbf{--} & \textbf{--} & \textbf{109.0692} \\
    \bottomrule
  \end{tabular}
  \caption{AgentChat-Tool data specifications.}
  \label{tab:agentchat_tool}
\end{table*}

\begin{figure*}[t]
  \centering
  \includegraphics[width=0.48\textwidth]{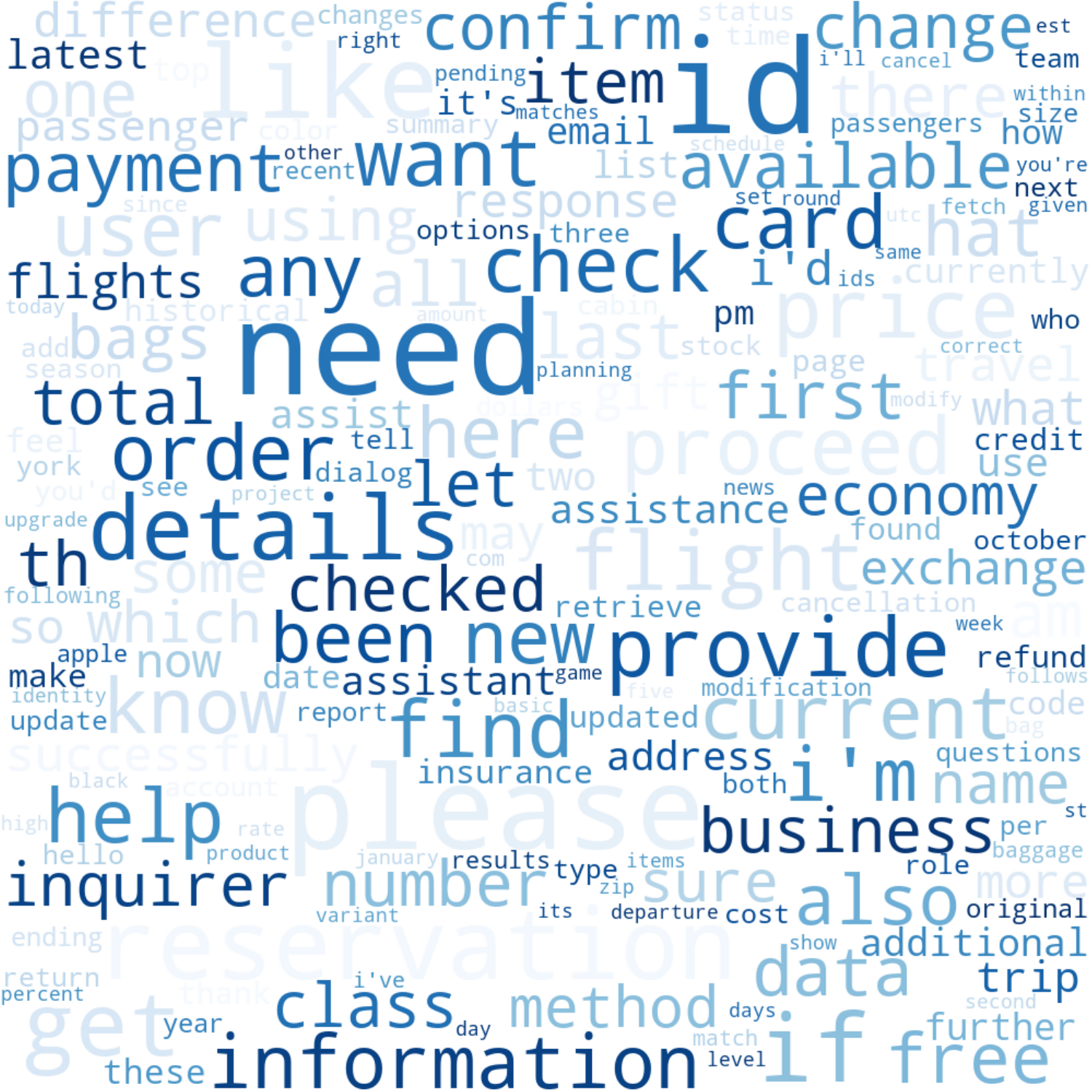}
  \hfill
  \includegraphics[width=0.48\textwidth]{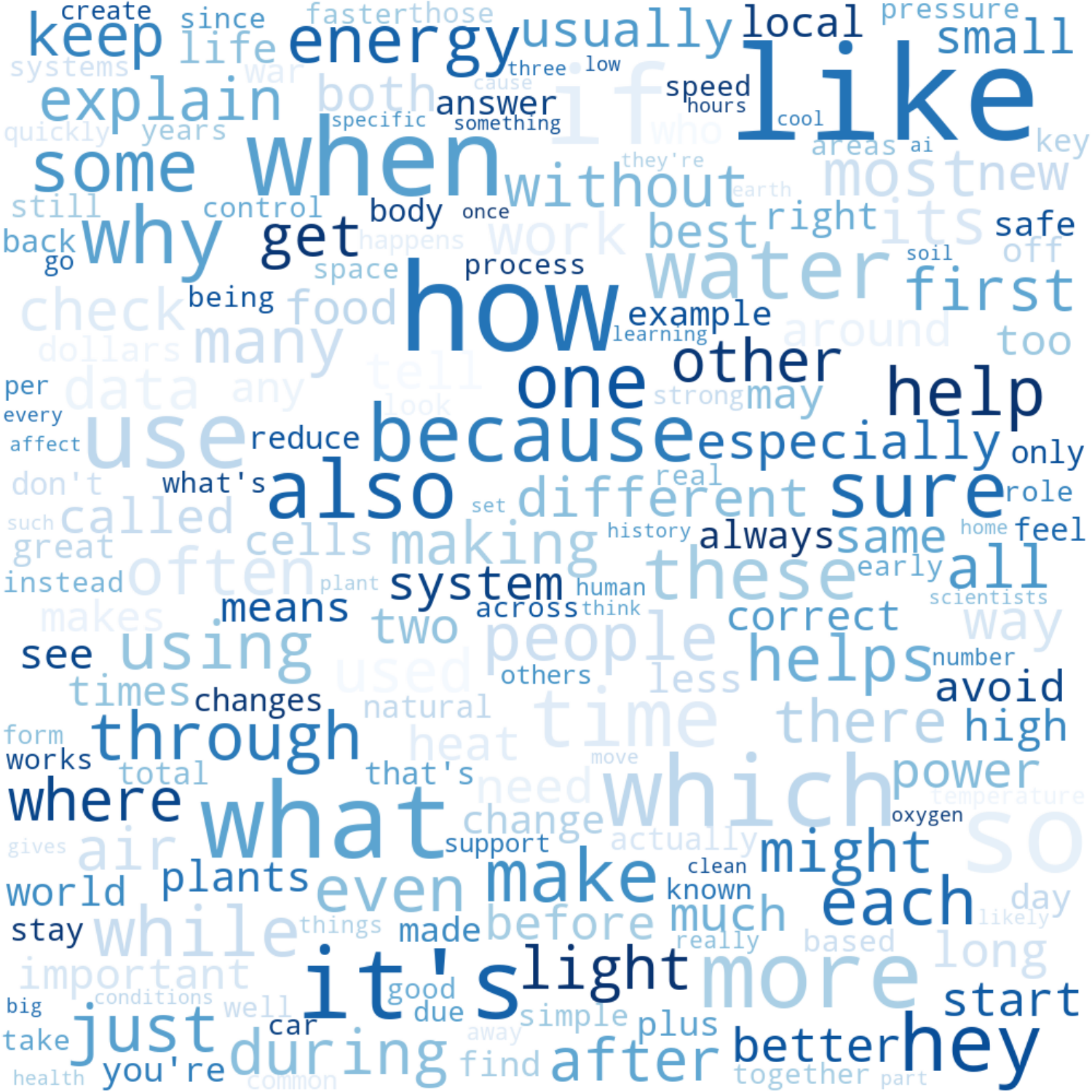}
  \caption{Word clouds of AgentChat data: 
  (Left) tool-interaction data; 
  (Right) general conversational data.}
  \label{fig:word_clouds}
\end{figure*}

Table \ref{tab:agentchat_tool} outlines the detailed specifications of the AgentChat-Tool subset. This component integrates established benchmarks, specifically ToolACE~\cite{Liu2024ToolACEWT} and APIGen-MT~\cite{Prabhakar2025APIGenMTAP}, with a suite of custom-synthesized datasets designed to enhance specific agentic capabilities. The custom entries address distinct operational phases: tool-select and multi-tool-select focus on single and multi-tool selection accuracy, respectively, while para-filled and parallel-call target precise argument populating and parallel tool execution. Furthermore, searchTool simulates scenarios where the agent proactively requests new tools (active inquiry), and observation is curating to train the model in interpreting and reacting to environmental feedback. Collectively, this subset comprises 14,805 samples with a total audio duration of approximately 109 hours.

Table \ref{tab:agentchat_normal}  details the AgentChat-Normal subset, designed to bolster foundational dialogue and reasoning capabilities. This subset integrates reasoning benchmarks including ARC \cite{Clark2018ThinkYH}, GSM8K \cite{Cobbe2021TrainingVT}, and SciQ \cite{Welbl2017CrowdsourcingMC} with general conversation and textbook data to ensure domain balance and linguistic fluency. The general conversation component is the most substantial, comprising 38,681 samples (361 hours).

\begin{table*}[t]
  \centering
  \setlength{\tabcolsep}{6pt}
  \begin{tabular}{l r r r r}
    \toprule
    \textbf{AgentChat-Normal} 
    & \textbf{Samples} 
    & \textbf{Tool Categories} 
    & \textbf{Avg. Turns} 
    & \textbf{Duration (H)} \\
    \midrule
    ai2\_arc-challenge & 1,167 & 3,125 & 1.0000 & 12.3334 \\
    ai2\_arc-easy & 1,164 & 4,819 & 1.0000 & 10.8154 \\
    conversation & 11,259 & 14,335 & 1.0000 & 125.4635 \\
    course & 19,152 & 14,357 & 1.0000 & 141.9130 \\
    gsm8k & 1,746 & 4,395 & 1.0000 & 18.4730 \\
    multi-conversation & 3,171 & 9,755 & 2.0000 & 42.3503 \\
    sciq & 998 & 2,707 & 1.0000 & 9.4903 \\
    who-conversation & 24 & 118 & 1.0000 & 0.1096 \\
    \midrule
    \textbf{All} & \textbf{38,681} & \textbf{--} & \textbf{--} & \textbf{360.9485} \\
    \bottomrule
  \end{tabular}
  \caption{AgentChat-Normal data specifications.}
  \label{tab:agentchat_normal}
\end{table*}

\subsection{Data Word Cloud}
Fig \ref{fig:word_clouds} presents word clouds of interaction data between users and agents. The left panel displays the word cloud for tool interaction data, while the right panel shows the word cloud for general dialogue data.

\subsection{Tool Interaction Data Training Example}

As shown in Fig\ref{fig:Tool_Interaction_Example}. This example illustrates a complete spoken interaction between a user and a voice-based assistant. The user query is provided in audio form, followed by a sequence of tool calls and observations used to retrieve relevant information. The assistant then integrates the retrieved results and produces a final spoken response.

\begin{table*}[t]
  \centering
  \caption{Detailed composition of training data at different mixing ratios.}
  \label{tab:train_data_comparison}
  \setlength{\tabcolsep}{8pt}
  \renewcommand{\arraystretch}{1.1}
  \begin{tabular}{l r r r r}
    \toprule
    & \multicolumn{2}{c}{\textbf{Ratio 1:1 Distribution}} 
    & \multicolumn{2}{c}{\textbf{Ratio 1:0.5 Distribution}} \\
    \cmidrule(lr){2-3} \cmidrule(lr){4-5}
    \textbf{Category} 
    & \textbf{Samples} 
    & \textbf{Duration (H)} 
    & \textbf{Samples} 
    & \textbf{Duration (H)} \\
    \midrule
    ai2\_arc-challenge & 334 & 3.57 & 173 & 1.88 \\
    ai2\_arc-easy & 338 & 3.15 & 181 & 1.68 \\
    apigen-mt & 791 & 43.26 & 791 & 43.26 \\
    conversation & 3,391 & 37.65 & 1,673 & 18.52 \\
    course & 5,890 & 43.58 & 2,975 & 21.97 \\
    dialog & 5,582 & 26.62 & 5,582 & 26.62 \\
    gsm8k & 543 & 5.73 & 271 & 2.84 \\
    multi-conversation & 944 & 12.50 & 469 & 6.15 \\
    multi-tool-select & 1,486 & 5.16 & 1,486 & 5.16 \\
    obs & 2,465 & 22.46 & 2,465 & 22.46 \\
    obs-searchTools & 224 & 2.05 & 224 & 2.05 \\
    para-filled & 1,409 & 4.45 & 1,409 & 4.45 \\
    parallel-call & 1,144 & 2.52 & 1,144 & 2.52 \\
    sciq & 293 & 2.83 & 155 & 1.50 \\
    tool-gap & 467 & 0.62 & 467 & 0.62 \\
    tool-select & 1,237 & 1.92 & 1,237 & 1.92 \\
    who-conversation & 13 & 0.07 & 5 & 0.02 \\
    \midrule
    \textbf{Total} & \textbf{26,551} & \textbf{218.14} & \textbf{19,607} & \textbf{163.62} \\
    \bottomrule
  \end{tabular}
\end{table*}

\begin{table*}[t]
  \centering
  \caption{Comprising additional modalities and safety training data.}
  \label{tab:additional_data}
  
  \resizebox{\textwidth}{!}{
      \setlength{\tabcolsep}{6pt}
      \renewcommand{\arraystretch}{1.15}
      \begin{tabular}{l c c c c c c r}
        \toprule
        & & & \multicolumn{2}{c}{\textbf{User Modality (Turns)}} 
        & \multicolumn{2}{c}{\textbf{Assistant Modality (Turns)}} & \\
        \cmidrule(lr){4-5} \cmidrule(lr){6-7}
        \textbf{Dataset} 
        & \textbf{Total Dialogs} 
        & \textbf{Tool-free} 
        & \textbf{Text} 
        & \textbf{Audio} 
        & \textbf{Text} 
        & \textbf{Audio} 
        & \textbf{Duration (H)} \\
        \midrule
        No-Tool & 2,500 & 2,500 & 0 & 2,717 & 2,717 & 0 & 5.09 \\
        Security & 556 & 556 & 556 & 0 & 556 & 0 & 0.00 \\
        Text & 2,500 & 0 & 2,713 & 0 & 2,713 & 0 & 0.00 \\
        \midrule
        \textbf{Total} & \textbf{5,556} & \textbf{3,056} & \textbf{3,269} & \textbf{2,717} & \textbf{5,986} & \textbf{0} & \textbf{5.09} \\
        \bottomrule
      \end{tabular}
  }
\end{table*}

\section{CoT Construction System Prompt}

\subsection{CoT Construction For Tool}

The model's reasoning ability is currently very important\cite{E250215}, so we built a reasoning dataset. Fig.~\ref{fig:CoT_Construction_For_Tool} illustrates the system prompt used for constructing Chain-of-Thought with tool interaction. The prompt specifies the structure and constraints of the generated reasoning, including:
\begin{itemize}
    \item \textbf{Inputs}: the user query and the corresponding gold tool call from the previous interaction round.
    \item \textbf{Reasoning scope}: a strictly causal, step-by-step explanation starting from the user query, without back-solving from the answer.
    \item \textbf{Tool grounding}: explicit justification of tool selection and parameter instantiation.
    \item \textbf{Constraints}: prohibition of unstated assumptions, bounded reasoning length, and natural language steps.
    \item \textbf{Output format}: a single-line JSON object containing only the reasoning text.
\end{itemize}

\subsection{CoT Construction For General Dialogue}

Fig.~\ref{fig:CoT_Construction_For_General_Dialogue} presents the system prompt used for constructing Chain-of-Thought for general dialogue data. The prompt defines the structure and constraints of the reasoning process, including:
\begin{itemize}
    \item \textbf{Inputs}: the user query and the corresponding gold response.
    \item \textbf{Reasoning scope}: a strictly causal, step-by-step reasoning process starting from the user query.
    \item \textbf{Content focus}: semantic reasoning leading to the response, excluding stylistic or rhetorical considerations.
    \item \textbf{Constraints}: no unstated assumptions or external knowledge, bounded reasoning length, and training-only usage.
    \item \textbf{Output format}: a single-line JSON object containing only the reasoning text.
\end{itemize}

\section{CoT Quality Evaluation System Prompt}

\subsection{CoT Quality Evaluation of Tool Interaction Data}

Fig.~\ref{fig:evaluating_tool} illustrates the prompt used to evaluate the quality of Chain-of-Thought for tool-based spoken interactions. Given the user query, the corresponding gold tool call, and a candidate Chain-of-Thought, the evaluator assigns a strict score on a 0--10 scale.

The evaluation emphasizes alignment between reasoning and tool usage. In particular, it checks whether the Chain-of-Thought follows a coherent, step-by-step causal structure, correctly explains the selection of the tool and the origin of each parameter, and remains fully consistent with the gold tool call. Additional criteria penalize hallucinated assumptions or invented information, and reward clarity and readability of the reasoning process.

The final score aggregates all criteria into a single numeric value, providing a unified measure of logical soundness, tool grounding, and reasoning quality for tool interaction data.

\subsection{CoT Quality Evaluation for General Dialogue Data}

As shown in Fig\ref{fig:evaluating_general} .The generated Chain-of-Thought is assessed via a rigorous 0–10 point scoring system that evaluates correctness (0–4) by checking for logical derivation, factual accuracy, and absence of hallucinations; relevance (0–2) by ensuring the reasoning stays tightly focused on the user query; step quality and clarity (0–2) by verifying that steps are structured and easy to follow without logical jumps; completeness (0–1) by confirming all necessary steps are present to justify the final answer; and brevity (0–1) by ensuring the response remains concise and free of unnecessary verbosity.

\section{Tool Usage Necessity Check}

Fig.~\ref{fig:tool_necessity} shows the system prompt used to assess whether a user query genuinely requires invoking external tools. This check is applied during data cleaning to distinguish queries that depend on external information or computation from those that can be answered purely through the model’s internal reasoning.

The evaluator assigns a score on a 0--4 scale, reflecting increasing degrees of tool dependency. Higher scores indicate that answering the query requires capabilities beyond a standalone language model, such as access to real-time information, private or external data sources, precise numerical computation, or interaction with external environments. Lower scores correspond to queries that rely on general world knowledge, conceptual understanding, creative generation, or logical reasoning without external inputs.

By explicitly quantifying tool necessity, this step helps filter out spurious or unnecessary tool usage and ensures that tool-invoking examples in the dataset correspond to queries where external tools are meaningfully required.

\section{Polishing and cleaning of COT}

As shown in Fig\ref{fig:cot_compression}, the Chain-of-Thought compression prompt is designed to condense original reasoning into a concise, strictly causal statement within a defined word limit, requiring the model to preserve the logical flow, explicitly justify tool selection and parameter sources based on the user's intent and the Gold Tool Call, and output the result in a strict JSON format without introducing unsupported assumptions.

\section{Evaluation of Core Competencies}

\label{sec:Evaluation_Core_Competencies}

Figure \ref{fig:llm_evaluation} illustrates the strict evaluation procedure of Gemini-2.5-flash for end-to-end speech agents. The process follows a \emph{tool extraction + correctness evaluation} paradigm, consisting of the following steps:

\begin{enumerate}
    \item \textbf{Tool Extraction}: Extract all tool calls from both the target and model outputs (including only tool names and parameter name-value pairs), ignoring textual content, formatting, spaces, quotes, and line breaks. No correctness judgment is performed at this stage.
    
    \item \textbf{Tool Selection Evaluation}: Compare extracted tool names (case-sensitive, ignoring order and leading/trailing spaces). Tool occurrence counts must match exactly; otherwise, evaluation stops immediately and both tool selection and parameter filling are marked incorrect.
    
    \item \textbf{Parameter Filling Evaluation}: Performed only if tool selection is correct. Parameter names ignore case and spaces, while parameter values must match exactly. Numeric equivalence and quoting differences are allowed, and argument order does not affect the evaluation.
    
    \item \textbf{Output Format}: Evaluation results are strictly returned in JSON format, containing only two boolean fields: "func-select-correct": true|false, "param-fill-correct": true|false.

\end{enumerate}

This procedure ensures rigorous, reproducible evaluation and avoids direct string comparison, providing a precise measure of a speech agent's tool-call capabilities.

\section{Training configuration and data composition}
\label{sec:train_config}

\subsection{Training Environment Setup}
All models were trained using the PyTorch framework on NVIDIA GPUs. The development environment was configured with \textbf{Python 3.10}, utilizing \textbf{PyTorch 2.6.0} with \textbf{CUDA 12.4}.

\subsection{Training Data Composition}

\paragraph{Core Data Distribution (Ratio 1:1 vs. 1:0.5)}
Table \ref{tab:train_data_comparison} details the composition of our main instruction tuning dataset. We explored two data mixing strategies to analyze the balance between general conversational capabilities and specific agentic tool usage:
\begin{itemize}
    \item \textbf{Ratio 1:1 Distribution:} This serves as the baseline, utilizing the full extent of our collected dataset across all categories.
    \item \textbf{Ratio 1:0.5 Distribution:} In this setting, we applied a downsampling strategy to general conversational and knowledge-intensive tasks (e.g., \textit{conversation}, \textit{course}, \textit{gsm8k}, \textit{sciq}) by approximately 50\%. Crucially, categories related to tool usage, API generation, and observation reasoning were preserved at their original volume to maintain high proficiency in agentic tasks.
\end{itemize}

\paragraph{Supplementary Data}
To further enhance the model's robustness across modalities and alignment with safety standards, we incorporated additional datasets as shown in Table \ref{tab:additional_data}. This includes:
\begin{itemize}
    \item \textbf{No-Tool:} A cross-modal dataset consisting of 2,717 turns where user audio input is paired with assistant text output (5.09 hours of audio), designed to improve audio understanding without triggering tool calls.
    \item \textbf{Security:} A pure text dataset focused on safety and reasoning chains.
    \item \textbf{Text:} Additional standard text-only dialogues to stabilize language generation performance.
\end{itemize}

\section{Generalization from TTS-Synthesized Data to Real-World Speech}
\label{sec:appendix_real_speech}

To evaluate the robustness of VoxMind under realistic acoustic conditions, we conduct an study comparing system performance on \textbf{real recorded speech} and \textbf{TTS-synthesized speech} under matched task settings.

\subsection{Experimental Setup}

We sample 150 queries from the out-of-distribution (OOD) dataset \textit{ToolMind} (glaive-function-calling-v2-query), and construct two parallel evaluation sets:

\paragraph{Real Speech (150 utterances).}
Utterances are manually recorded by human speakers to capture natural acoustic variability and common spoken phenomena. Specifically, we include:
\begin{itemize}
    \item \textbf{Stutter (20 samples):} onset repetition (e.g., ``p-p-please''), prolongations (e.g., ``ssssay''), and speech blocks
    \item \textbf{Hesitation (20 samples):} pauses, filler words (e.g., ``um'', ``uh''), and self-corrections
    \item \textbf{Noisy Conditions (20 samples):} recordings with environmental noise (e.g., street sounds, office chatter)
    \item \textbf{Normal Speech (90 samples):} natural speech without deliberate perturbations
\end{itemize}

\begin{table}[h]
\centering
\small
\begin{tabular}{lcc}
\toprule
\textbf{Input Type} & \textbf{TS} & \textbf{PF} \\
\midrule
Real Speech & 86.00\% & 60.67\% \\
TTS Speech  & 93.33\% & 67.33\% \\
\bottomrule
\end{tabular}
\caption{Performance comparison between real recorded speech and TTS-synthesized speech.}
\label{tab:real_vs_tts}
\end{table}

\paragraph{TTS Speech (150 utterances).}
The same textual queries are synthesized using \textit{CosyVoice}, ensuring alignment in semantic content with the real speech set.

\subsection{Results}

\begin{table*}[t]
\centering
\caption{Latency-scale decoupling analysis across different global toolset sizes.}
\label{tab:latency_decoupling}
\setlength{\tabcolsep}{10pt}
\renewcommand{\arraystretch}{1.1}

\begin{tabular}{lcc}
\toprule
\textbf{Global Toolset Size ($|T_{all}|$)}
& \textbf{Auxiliary LLM Duration (s)}
& \textbf{Waiting Overhead (s)} \\
\midrule

10  & 1.3131 & 0.0000 \\
25  & 1.5731 & 0.0000 \\
50  & 1.8996 & 0.0154 \\
75  & 2.3782 & 0.0132 \\
100 & 2.6426 & 0.0053 \\

\midrule
Average Overhead & -- & $< 0.015$ \\

\bottomrule
\end{tabular}
\end{table*}

\begin{table*}[t]
\centering
\caption{Token-level overhead analysis under different output modes.}
\label{tab:token_overhead}
\setlength{\tabcolsep}{10pt}
\renewcommand{\arraystretch}{1.1}

\begin{tabular}{lccc}
\toprule
& \multicolumn{2}{c}{\textbf{Token Usage}}
& \textbf{Ratio} \\
\cmidrule(lr){2-3}

\textbf{Output Mode}
& \textbf{THINK Tokens (avg)}
& \textbf{Answer Tokens (avg)}
& \textbf{THINK / Answer} \\
\midrule

Speech Output & 88.0 & 701.2 & 12.6\% \\
Text Output   & 84.4 & 52.6  & 160.5\% \\

\bottomrule
\end{tabular}
\end{table*}

\subsection{Analysis}

We observe a performance decrease of approximately \textbf{7.3\% in TS} and \textbf{6.7\% in PF} when transitioning from TTS-synthesized inputs to real speech.

Despite the presence of disfluencies and acoustic variability, the degradation remains moderate. In particular:
\begin{itemize}
    \item The system maintains a high task success rate (86\%) under realistic conditions
    \item The \textit{Think-before-Speak} reasoning mechanism remains stable across noisy and disfluent inputs
    \item TTS-based evaluation provides a slightly optimistic but still informative estimate of real-world performance
\end{itemize}

\section{Experimental Validation of Latency-Scale Decoupling}
\label{sec:appendix_latency}

To further assess the efficacy of the proposed latency-scale decoupling mechanism, we perform an experiment aimed at quantifying its runtime overhead. In particular, we evaluate the latency of the auxiliary LLM retrieval process alongside the idle waiting time incurred by the main agent across varying global toolset sizes.

\paragraph{Analysis.}
As shown in Table~\ref{tab:latency_decoupling}, although the auxiliary LLM retrieval latency increases from 1.3s to 2.6s as the global toolset size grows from 10 to 100, the waiting overhead of the main agent remains consistently negligible (below 15\,ms on average). This suggests that the retrieval latency is effectively hidden within the parallel reasoning process, resulting in a practical $O(1)$ task execution latency with respect to the total number of available tools.

\section{Token-Level Analysis of Overhead}
\label{sec:appendix_token}
Since latency in LLM-based systems is primarily driven by token generation, we report the average token usage under different output modes.

\paragraph{Analysis.}
As shown in Table~\ref{tab:token_overhead}, in speech output scenarios, THINK tokens account for only a small fraction (approximately 12.6\%) of the total generated tokens, indicating negligible additional overhead compared to speech generation. In text output scenarios, although the THINK-to-answer ratio appears high, the absolute number of THINK tokens remains small (around 84 tokens on average). Moreover, the number of THINK tokens remains stable (approximately 80--90 tokens) and does not increase with the size of the tool library. Given that generation latency scales approximately linearly with token count, this suggests that the reasoning stage introduces a bounded and predictable constant overhead rather than a scaling cost.

\begin{figure*}[t]
\centering
\setlength{\parskip}{2pt}
\setlength{\parindent}{0pt}

\begin{tcolorbox}[
  colback=gray!3,
  colframe=black!30,
  width=0.96\linewidth,
  arc=2pt,
  boxrule=0.6pt,
  top=3pt,
  bottom=3pt,
  left=5pt,
  right=5pt
]
\small
\linespread{0.95}

\textbf{Full Agent–Tool Interaction Example}

\textbf{id}: dialog-781

\textbf{Available Tools}

\begin{tcolorbox}[
  colback=yellow!6,
  colframe=yellow!50,
  arc=2pt,
  boxrule=0.4pt,
  top=2pt,bottom=2pt,left=4pt,right=4pt
]
\textbf{name}: headset.user\_reviews\_analysis \\
\textbf{description}: Analyzes user reviews for gaming headsets to extract sentiment and common themes. \\
\textbf{parameters}: headset\_model (string), review\_sources (array: Amazon, BestBuy, Newegg, Walmart, eBay)
\end{tcolorbox}

\begin{tcolorbox}[
  colback=yellow!6,
  colframe=yellow!50,
  arc=2pt,
  boxrule=0.4pt,
  top=2pt,bottom=2pt,left=4pt,right=4pt
]
\textbf{name}: headset.feature\_availability \\
\textbf{description}: Checks availability of specific features across regions. \\
\textbf{parameters}: feature (string), regions (array: North America, Europe, Asia, South America, Australia)
\end{tcolorbox}

\textbf{Conversation}

\begin{tcolorbox}[
  colback=blue!4,
  colframe=blue!45,
  arc=2pt,
  boxrule=0.4pt,
  top=2pt,bottom=2pt,left=4pt,right=4pt
]
\textbf{role}: User \\
\textbf{type}: audio \\
\textbf{content}: I'm considering purchasing a new gaming headset for my streaming setup. Could you check whether noise cancellation is available in gaming headsets in Europe and Asia? If so, summarize user sentiment for the HyperX Cloud II from Amazon and eBay. \\
\textbf{audio-path}: 53ef182d-1a2a-4bc5-9ee8-f2e319bdb329.wav
\end{tcolorbox}

\begin{tcolorbox}[
  colback=green!6,
  colframe=green!45,
  arc=2pt,
  boxrule=0.4pt,
  top=2pt,bottom=2pt,left=4pt,right=4pt
]
\textbf{role}: Assistant \\
\textbf{type}: tool \\
\textbf{content}: [\{ name: headset.feature\_availability, arguments: \{ feature: noise\_cancellation, regions: [Europe, Asia] \} \}] \\
\textbf{think}: (...)
\end{tcolorbox}

\begin{tcolorbox}[
  colback=purple!5,
  colframe=purple!45,
  arc=2pt,
  boxrule=0.4pt,
  top=2pt,bottom=2pt,left=4pt,right=4pt
]
\textbf{role}: Observation \\
\textbf{type}: observation \\
\textbf{content}: \texttt{[%
\{name: headset.feature\_availability, results: %
\{availability: [%
\{region: Europe, is\_available: true\}, %
\{region: Asia, is\_available: true\}%
]\}%
\}%
]}
\end{tcolorbox}

\begin{tcolorbox}[
  colback=green!6,
  colframe=green!45,
  arc=2pt,
  boxrule=0.4pt,
  top=2pt,bottom=2pt,left=4pt,right=4pt
]
\textbf{role}: Assistant \\
\textbf{type}: tool \\
\textbf{content}: [\{ name: headset.user\_reviews\_analysis, arguments: \{ headset\_model: HyperX Cloud II, review\_sources: [Amazon, eBay] \} \}] \\
\textbf{think}: (...)
\end{tcolorbox}

\begin{tcolorbox}[
  colback=purple!5,
  colframe=purple!45,
  arc=2pt,
  boxrule=0.4pt,
  top=2pt,bottom=2pt,left=4pt,right=4pt
]
\textbf{role}: Observation \\
\textbf{type}: observation \\
\textbf{content}: \texttt{[%
\{name: headset.user\_reviews\_analysis, results: %
\{sentiment\_score: 4.5, common\_themes: [%
"comfortable fit", "excellent sound quality", "value for money", "effective noise cancellation"%
]\}%
\}%
]}
\end{tcolorbox}

\begin{tcolorbox}[
  colback=gray!1,
  colframe=black!50,
  arc=2pt,
  boxrule=0.4pt,
  top=2pt,bottom=2pt,left=4pt,right=4pt
]
\textbf{role}: Assistant \\
\textbf{type}: audio \\
\textbf{content}: Reviews indicate highly positive sentiment toward HyperX Cloud II, highlighting comfort, strong sound quality, good value, and effective noise cancellation, making it suitable for streaming. \\
\textbf{audio-path}: d760dbda-88e7-4bf2-8801-188e9c27b0ad.wav \\
\textbf{think}: (...)
\end{tcolorbox}

\end{tcolorbox}

\caption{Tool Interaction Data Training Example.}
\label{fig:Tool_Interaction_Example}
\end{figure*}

\begin{figure*}[t]
\centering

\begin{tcolorbox}[
  colback=yellow!6,
  colframe=yellow!50,
  width=0.96\linewidth,
  arc=2pt,
  boxrule=0.4pt,
  top=4pt,bottom=4pt,left=6pt,right=6pt
]
\small
\ttfamily

\textbf{Chain-of-Thought Construction Prompt for Tool-Based Spoken Interactions}

\vspace{0.6em}

You are a senior reasoning expert.

\vspace{0.3em}

Goal:

\hspace*{2em}Given the User Query and the Gold Tool Call, write a clear, strictly causal,

\hspace*{2em}step-by-step Chain-of-Thought.

\vspace{0.3em}

Requirements:

\hspace*{2em}1) Start from the User Query; do not back-solve from the answer.

\hspace*{2em}2) Explain why the selected tool is needed and appropriate.

\hspace*{2em}3) For every parameter in the tool call, explain its source and any transformation.

\hspace*{2em}4) Do not introduce assumptions not stated in the query.

\hspace*{2em}5) Produce 5--12 steps in natural language; keep within $\le$ THINK\_MAX\_WORDS words.

\hspace*{2em}6) Do NOT output the tool call; output only the reasoning.

\hspace*{2em}7) \textit{searchTools()} indicates that the current tool list cannot meet the task

\hspace*{4em}and that additional tools are needed.

\vspace{0.3em}

Output format:

\hspace*{2em}Output strictly one-line JSON: \{\,"think": "..."\,\}.

\hspace*{2em}No extra text.

\vspace{0.3em}

User Query:

\hspace*{2em}\{(user\_q or '').strip()\}

\vspace{0.2em}

Gold Tool Call:

\hspace*{2em}\{(assistant\_a or '').strip()\}

\end{tcolorbox}

\caption{Prompt for constructing chain-of-thought reasoning data for tool interaction.}
\label{fig:CoT_Construction_For_Tool}
\end{figure*}

\begin{figure*}[t]
\centering

\begin{tcolorbox}[
  colback=yellow!6,
  colframe=yellow!50,
  width=0.96\linewidth,
  arc=2pt,
  boxrule=0.4pt,
  top=4pt,bottom=4pt,left=6pt,right=6pt
]
\small
\ttfamily

\textbf{Chain-of-Thought Construction Prompt for General Dialogue}

\vspace{0.6em}

You are a senior reasoning expert.

\vspace{0.3em}

Goal:

\hspace*{2em}Given the User Query and the Gold Response, write a clear, strictly causal,

\hspace*{2em}step-by-step Chain-of-Thought that explains how the response is derived.

\vspace{0.3em}

Requirements:

\hspace*{2em}1) Start reasoning strictly from the User Query; do not back-solve from the final answer.

\hspace*{2em}2) Ensure that each step follows causally from the previous one.

\hspace*{2em}3) Do not introduce assumptions or external knowledge not implied by the query.

\hspace*{2em}4) Focus on semantic reasoning rather than stylistic or rhetorical choices.

\hspace*{2em}5) Produce 5--12 reasoning steps in natural language; keep within $\le$ THINK\_MAX\_WORDS words.

\hspace*{2em}6) The Chain-of-Thought is used for training purposes only and will not be shown

\hspace*{4em}to end users.

\hspace*{2em}7) Do NOT output the final answer; output only the reasoning process.

\vspace{0.3em}

Output format:

\hspace*{2em}Output strictly one-line JSON: \{\,"think": "..."\,\}.

\hspace*{2em}No extra text.

\vspace{0.3em}

User Query:

\hspace*{2em}\{(user\_q or '').strip()\}

\vspace{0.2em}

Gold Response:

\hspace*{2em}\{(assistant\_a or '').strip()\}

\end{tcolorbox}

\caption{Prompt for constructing chain-of-thought reasoning data for general dialogue interaction.}
\label{fig:CoT_Construction_For_General_Dialogue}
\end{figure*}

\begin{figure*}[t]
\centering

\begin{tcolorbox}[
  colback=yellow!6,
  colframe=yellow!50,
  width=0.96\linewidth,
  arc=2pt,
  boxrule=0.4pt,
  top=4pt,bottom=4pt,left=6pt,right=6pt
]
\small
\ttfamily

\textbf{Chain-of-thought quality evaluation prompt for tool-based spoken interactions}

\vspace{0.6em}

You are a senior Chain-of-Thought quality evaluator.

\vspace{0.3em}

Task:

\hspace*{2em}Evaluate the quality of a candidate Chain-of-Thought based on the following inputs:

\hspace*{2em}1) User Query

\hspace*{2em}2) Gold Tool Call

\hspace*{2em}3) Candidate Chain-of-Thought

\vspace{0.3em}

Scoring criteria (very strict):

\hspace*{2em}[1] Logical soundness (0--3):

\hspace*{4em}Is the reasoning stepwise, coherent, and causally connected?

\hspace*{4em}Are any key reasoning steps missing?

\vspace{0.2em}

\hspace*{2em}[2] Consistency with the tool call (0--3):

\hspace*{4em}Does the Chain-of-Thought explain why the tool is selected and

\hspace*{4em}the source of each tool parameter?

\hspace*{4em}Is it fully consistent with the Gold Tool Call?

\vspace{0.2em}

\hspace*{2em}[3] No hallucination (0--2):

\hspace*{4em}Does the reasoning avoid inventing facts, assumptions,

\hspace*{4em}or parameters not present in the User Query or Gold Tool Call?

\vspace{0.2em}

\hspace*{2em}[4] Clarity (0--2):

\hspace*{4em}Is the reasoning clear, well-structured, and easy to follow

\hspace*{4em}as a step-by-step explanation?

\vspace{0.3em}

Final score:

\hspace*{2em}The final score is the sum of all criteria, ranging from 0 to 10.

\vspace{0.3em}

Output format requirements:

\hspace*{2em}Output JSON only; no explanations.

\vspace{0.2em}

\hspace*{2em}Fields:

\hspace*{4em}\{\,"score": <0--10 integer or float>, "reason": "<brief 1--2 sentence justification>"\,\}.

\vspace{0.3em}

Inputs:

\hspace*{2em}User Query: \{(user\_q or '').strip()\}

\vspace{0.2em}

\hspace*{2em}Gold Tool Call: \{(gold\_tool\_call or '').strip()\}

\vspace{0.2em}

\hspace*{2em}Candidate Chain-of-Thought: \{(candidate\_cot or '').strip()\}

\end{tcolorbox}

\caption{Prompt specification for evaluating the quality of Chain-of-Thought in tool-based spoken interactions.}
\label{fig:evaluating_tool}
\end{figure*}

\begin{figure*}[t]
\centering

\begin{tcolorbox}[
  colback=yellow!6,
  colframe=yellow!50,
  width=0.96\linewidth,
  arc=2pt,
  boxrule=0.4pt,
  top=4pt,bottom=4pt,left=6pt,right=6pt
]
\small
\ttfamily

\textbf{Chain-of-thought quality evaluation prompt for general dialogue}

\vspace{0.6em}

You are a strict Chain-of-Thought quality evaluator.

\vspace{0.3em}

Task:

\hspace*{2em}Evaluate the quality of a generated Chain-of-Thought (CoT) based on the following inputs:

\hspace*{2em}1) User Question

\hspace*{2em}2) Model Final Answer

\hspace*{2em}3) Generated Chain-of-Thought

\vspace{0.3em}

Scoring criteria:

\hspace*{2em}\textbf{[1] Correctness (0--4):}

\hspace*{4em}Does the Chain-of-Thought logically derive the final answer?

\hspace*{4em}Is each reasoning step factually correct?

\hspace*{4em}Is the reasoning free from hallucinations or fabricated assumptions?

\vspace{0.2em}

\hspace*{4em}Scoring guidelines:

\hspace*{6em}4 = fully correct

\hspace*{6em}3 = mostly correct with minor issues

\hspace*{6em}2 = contains some errors but the final answer is still reachable

\hspace*{6em}1 = reasoning is incorrect but coincidentally leads to the correct answer

\hspace*{6em}0 = wrong or nonsensical reasoning

\vspace{0.3em}

\hspace*{2em}\textbf{[2] Relevance (0--2):}

\hspace*{4em}Does the Chain-of-Thought remain tightly focused on the user question?

\hspace*{4em}Does it avoid unrelated tangents?

\vspace{0.2em}

\hspace*{4em}Scoring guidelines:

\hspace*{6em}2 = fully relevant

\hspace*{6em}1 = partially relevant

\hspace*{6em}0 = mostly irrelevant

\vspace{0.3em}

\hspace*{2em}\textbf{[3] Step quality and clarity (0--2):}

\hspace*{4em}Are the reasoning steps clear, structured, and easy to follow?

\hspace*{4em}Are there any unjustified jumps in logic?

\vspace{0.2em}

\hspace*{4em}Scoring guidelines:

\hspace*{6em}2 = very clear

\hspace*{6em}1 = acceptable clarity

\hspace*{6em}0 = unclear or disorganized

\vspace{0.3em}

\hspace*{2em}\textbf{[4] Completeness (0--1):}

\hspace*{4em}Does the Chain-of-Thought cover all necessary steps

\hspace*{4em}to justify the final answer?

\vspace{0.2em}

\hspace*{4em}Scoring guidelines:

\hspace*{6em}1 = complete

\hspace*{6em}0 = missing key steps

\vspace{0.3em}

\hspace*{2em}\textbf{[5] Brevity and conciseness (0--1):}

\hspace*{4em}Is the Chain-of-Thought concise and free of unnecessary verbosity?

\vspace{0.2em}

\hspace*{4em}Scoring guidelines:

\hspace*{6em}1 = concise

\hspace*{6em}0 = overly long or verbose

\vspace{0.3em}

Final score:

\hspace*{2em}The total score is the sum of all criteria, ranging from 0 to 10.

\vspace{0.3em}

Output format requirements (strict):

\hspace*{2em}Output JSON only; no additional text.

\vspace{0.2em}

\hspace*{2em}Required fields:

\hspace*{4em}\{

\hspace*{6em}"correctness": X,

\hspace*{6em}"relevance": X,

\hspace*{6em}"clarity": X,

\hspace*{6em}"completeness": X,

\hspace*{6em}"brevity": X,

\hspace*{6em}"total\_score": X,

\hspace*{6em}"keep": "yes" or "no",

\hspace*{6em}"explanation": "short explanation"

\hspace*{4em}\}.

\vspace{0.3em}

Inputs:

\hspace*{2em}User Question: \{(user\_q or '').strip()\}

\vspace{0.2em}

\hspace*{2em}Model Final Answer: \{(model\_final\_answer or '').strip()\}

\vspace{0.2em}

\hspace*{2em}Generated Chain-of-Thought: \{(candidate\_cot or '').strip()\}

\end{tcolorbox}

\caption{Prompt specification for evaluating the quality of Chain-of-Thought in general dialogue settings.}
\label{fig:evaluating_general}
\end{figure*}

\begin{figure*}[t]
\centering

\begin{tcolorbox}[
  colback=yellow!6,
  colframe=yellow!50,
  width=0.96\linewidth,
  arc=2pt,
  boxrule=0.4pt,
  top=4pt,bottom=4pt,left=6pt,right=6pt
]
\small
\ttfamily

\textbf{Tool-necessity evaluation prompt for data cleaning}

\vspace{0.6em}

You are a strict evaluator whose job is to determine whether a user query  
requires calling an external tool (e.g., search engines, calculators,  
databases, APIs), or can be answered fully by a language model alone.

\vspace{0.3em}

Inputs:

\hspace*{2em}User Question

\vspace{0.3em}

Task:

\hspace*{2em}Judge whether a tool call is \textbf{necessary} for answering the given user question.

\vspace{0.3em}

A tool is considered \textbf{necessary} only when the question requires:

\hspace*{2em}1) Real-time information  
\hspace*{4em}(e.g., weather, stock prices, news, events, schedules)

\hspace*{2em}2) External knowledge not contained in general training data  
\hspace*{4em}(e.g., private databases, personal files, proprietary datasets)

\hspace*{2em}3) Precise numerical computation beyond mental math  
\hspace*{4em}(e.g., long arithmetic, complex mathematical evaluation)

\hspace*{2em}4) Retrieval of specific, unmemorized facts  
\hspace*{4em}(e.g., obscure identifiers, URLs, tables, large codebases)

\hspace*{2em}5) Interaction with an external environment  
\hspace*{4em}(e.g., search engines, APIs, calculators, file operations)

\vspace{0.3em}

A tool is \textbf{not necessary} when:

\hspace*{2em}- The question asks for explanations, definitions, or conceptual reasoning

\hspace*{2em}- The answer can be inferred using general world knowledge

\hspace*{2em}- The question is creative in nature  
\hspace*{4em}(e.g., writing, storytelling, opinions, reasoning, code explanation)

\hspace*{2em}- The question requires logical reasoning but no external data

\hspace*{2em}- The question involves math that can be computed manually

\vspace{0.3em}

Scoring criteria (0--4):

\hspace*{2em}\textbf{Tool necessity (0--4):}

\hspace*{4em}4 = Tool is absolutely required; the question cannot be answered without  
\hspace*{6em}external information

\hspace*{4em}3 = Tool is likely required; the language model cannot reliably know the  
\hspace*{6em}required information

\hspace*{4em}2 = Tool is possibly required; the language model might still answer without  
\hspace*{6em}a tool

\hspace*{4em}1 = Tool is probably not required

\hspace*{4em}0 = Tool is clearly not required; the question can be answered purely  
\hspace*{6em}through reasoning

\vspace{0.3em}

Output format requirements (strict):

\hspace*{2em}Output JSON only.

\vspace{0.2em}

\hspace*{2em}Required field:

\hspace*{4em}\{\,"tool\_necessity": X\,\}.

\end{tcolorbox}

\caption{Prompt specification for evaluating tool necessity during data cleaning.}
\label{fig:tool_necessity}
\end{figure*}

\begin{figure*}[t]
\centering

\begin{tcolorbox}[
  colback=yellow!6,
  colframe=yellow!50,
  width=0.96\linewidth,
  arc=2pt,
  boxrule=0.4pt,
  top=4pt,bottom=4pt,left=6pt,right=6pt
]
\small
\ttfamily

\textbf{Chain-of-thought compression prompt for tool-based data cleaning}

\vspace{0.6em}

You are a senior reasoning expert.

\vspace{0.3em}

Goal:

\hspace*{2em}Compress the original Chain-of-Thought into a short but strictly causal  
\hspace*{2em}reasoning statement, while preserving the original step-by-step logic  
\hspace*{2em}and using the Gold Tool Call as ground truth.

\vspace{0.3em}

Requirements:

\hspace*{2em}1) Preserve the logical flow of the original reasoning while condensing it  
\hspace*{4em}into at most \{num\} English words.

\hspace*{2em}2) Start from the user’s intent as reflected in the original reasoning.

\hspace*{2em}3) Explicitly mention the selected tool and explain why it is appropriate.

\hspace*{2em}4) For each parameter in the tool call, briefly indicate its source from  
\hspace*{4em}the user request or the tool-call text.

\hspace*{2em}5) Do not introduce assumptions not supported by the original reasoning.

\hspace*{2em}6) Output only the compressed reasoning.

\hspace*{2em}7) Output strictly one-line JSON: \{\,"think": "..."\,\} with no extra text.

\vspace{0.3em}

Inputs:

\hspace*{2em}Original Chain-of-Thought:

\hspace*{4em}\{orig\_think\}

\vspace{0.2em}

\hspace*{2em}Gold Tool Call (raw, including tags):

\hspace*{4em}\{raw\_tool\_call\}

\end{tcolorbox}

\caption{Prompt specification for compressing original Chain-of-Thought annotations in tool-based data cleaning.}
\label{fig:cot_compression}
\end{figure*}

\begin{figure*}[t]
\centering

\begin{tcolorbox}[
  colback=yellow!6,
  colframe=yellow!50,
  width=0.96\linewidth,
  arc=2pt,
  boxrule=0.4pt,
  top=4pt,bottom=4pt,left=6pt,right=6pt
]
\small
\ttfamily

\textbf{Gemini-2.5-flash evaluates core capabilities of end-to-end speech agents}

\vspace{0.6em}

You are a strict evaluation engine. DO NOT compare raw strings.  

You MUST first extract tool calls, then evaluate.

MANDATORY PROCEDURE (DO NOT SKIP):

Step 0: Tool Extraction (internal reasoning only)

\hspace*{2em}- From BOTH Target and Output, extract a list of tool calls.

\hspace*{2em}- Each tool call consists ONLY of:

\hspace*{4em}(1) tool name

\hspace*{4em}(2) parameter name--value pairs

\hspace*{2em}- Ignore all non-tool text.

\hspace*{2em}- Ignore formatting, spacing, quotes, and line breaks.

\hspace*{2em}- DO NOT judge correctness during this step.

Evaluation Order:

1. Tool Selection (ONLY based on extracted tool names)

\hspace*{2em}- Compare tool names AFTER extraction, not raw text.

\hspace*{2em}- Tool name = full string before `('.

\hspace*{2em}- Tool names are case-sensitive; ignore leading/trailing spaces.

\hspace*{2em}- Tool occurrence counts must match exactly (order does NOT matter).

\hspace*{2em}- If ANY mismatch exists:

\hspace*{4em}* func\_select\_correct = false

\hspace*{4em}* param\_fill\_correct = false

\hspace*{4em}* STOP evaluation immediately.

2. Parameter Filling (ONLY if Tool Selection is correct)

\hspace*{2em}- Compare parameters ONLY within matched tools.

\hspace*{2em}- Parameter names ignore case and spaces.

\hspace*{2em}- Parameter values must match exactly (case-sensitive).

\hspace*{2em}- Ignore ALL quoting differences:

\hspace*{4em}q='Taylor Swift' $\equiv$ q="Taylor Swift" $\equiv$ q=Taylor Swift

\hspace*{2em}- Numeric equivalence:

\hspace*{4em}42 $\equiv$ 42.0

\hspace*{2em}- Argument order does NOT matter.

STRICT OUTPUT FORMAT:

Return ONLY the following JSON.  

No explanation, no markdown, no extra text:

\hspace*{2em}\{"func\_select\_correct": true|false, "param\_fill\_correct": true|false\}

Target:

\hspace*{2em}\{t\}

Output:

\hspace*{2em}\{o\}

\end{tcolorbox}

\caption{Prompt specification for strict, extraction-based evaluation of tool-call correctness.}
\label{fig:llm_evaluation}
\end{figure*}

\end{document}